\documentclass[aps,pre,amsmath,superscriptaddress,amssymb,twocolumn,floatfix]{revtex4}     
\usepackage{graphicx}
\usepackage{color}

\newcommand\x{0.35}
\newcommand\xx{0.171}
\newcommand*{\ehb}{\epsilon_{\rm HB}}
\newcommand*{\spin}{{\bf S}}
\newcommand*{\coex}{{\bf C}}
\newcommand*{\hh}{{\textendash}}

\begin{document}
\title{Revisiting Speedy's stability-limit conjecture for the behavior of supercooled water}
\title{Revisiting Speedy's stability-limit conjecture}
\title{The stability-limit conjecture revisited}

\author{Pheerawich Chitnelawong}
\affiliation{Department of Physics, St. Francis Xavier University, Antigonish, NS, B2G 2W5, Canada}

\author{Francesco Sciortino}
\affiliation{Dipartimento di Fisica,
Universit\`a di Roma {\em La Sapienza},
Piazzale A. Moro 5, 00185 Roma, Italy}

\author{Peter H. Poole}
\email[Corresponding author: ]{ppoole@stfx.ca}
\affiliation{Department of Physics, St. Francis Xavier University, Antigonish, NS, B2G 2W5, Canada}

\begin{abstract}
The stability-limit conjecture (SLC) proposes that the liquid spinodal of water returns to positive pressure in the supercooled region, and that the apparent divergence of water's thermodynamic response functions as temperature decreases are explained by the approach to this reentrant spinodal.
Subsequently, it has been argued that the predictions of the SLC are inconsistent with general thermodynamic principles.
Here we reconsider the thermodynamic viability of the SLC by examining a model equation of state for water first studied to clarify the relationship of the SLC to the proposed liquid-liquid phase transition in supercooled water.
By demonstrating that a binodal may terminate on a spinodal at a point that is not a critical point, we show  that the SLC is thermodynamically permissible in a system that has both a liquid-gas and a liquid-liquid phase transition.  We also describe and clarify other unusual thermodynamic behavior that may arise in such a system, particularly that associated with the so-called ``critical-point-free" scenario for a liquid-liquid phase transition, which may apply to the case of liquid Si.
\end{abstract}

\date{\today}
\maketitle

\section{introduction}

\begin{figure}[b]
\centerline{\includegraphics[scale=0.25]{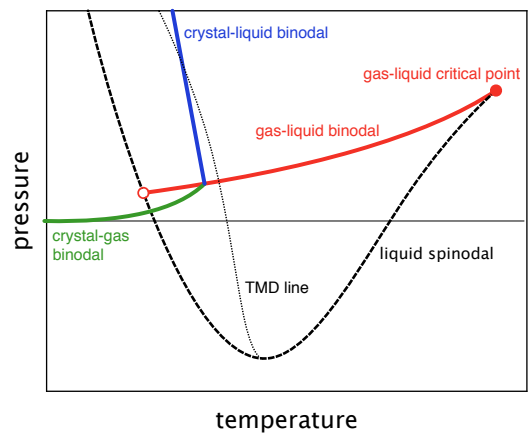}}
\caption{Schematic $T$-$P$ phase diagram for the SLC.  Gas-liquid, crystal-liquid, and crystal-gas binodals are shown as solid lines and meet at the triple point.  The gas-liquid critical point (filled circle), liquid spinodal (dashed), and TMD line (dotted) are also shown.  The open circle locates the intersection of the liquid spinodal and the metastable extension of the gas-liquid binodal.}
\label{slc}
\end{figure}

The ``stability-limit conjecture" (SLC) was one of the first attempts to provide a unified thermodynamic understanding of the multitude of anomalies of cold and supercooled water~\cite{Speedy:1976,Speedy:1982fd}.  
The predictions of the SLC are illustrated schematically in Fig.~\ref{slc}.  The principal claim of the SLC is the behavior that it proposes for the limit of metastability, or spinodal, of the liquid phase.  In the temperature-pressure ($T$-$P$) phase diagram, the liquid spinodal begins at the gas-liquid (G-L) critical point and lies below the G-L coexistence curve, or binodal.  In a simple liquid, the liquid spinodal is expected to decrease monotonically into the negative pressure region as $T$ decreases.  However, as first noted by Speedy~\cite{Speedy:1982fd}, the existence of a density maximum in liquid water can alter the behavior of the liquid spinodal.  Specifically, Speedy showed that if the temperature of maximum density (TMD) line intersects the liquid spinodal, then the slope of the spinodal in the $T$-$P$ plane will change sign, creating the potential for the spinodal to be reentrant, that is, return to positive $P$.  If a reentrant liquid spinodal occurs at positive $P$ in the supercooled liquid, then the rapid increase of thermodynamic response functions upon cooling can be attributed to the divergence of fluctuations expected at a spinodal.

The SLC stimulated a large body of work focussing on the overall thermodynamic properties of water 
and played a pivotal role in guiding the discovery of other scenarios.  The liquid-liquid critical point (LLCP) scenario
was proposed as a direct result of an attempt to test the predictions of the SLC in simulations of ST2 water~\cite{Poole:1992p5118,poole1993phase,sciortino1997line}.

In the LLCP scenario, the observed anomalies of water arise as a consequence of a line of first-order liquid-liquid phase transitions that occurs in the phase diagram for supercooled water, terminating at a critical point.  Below the temperature of this critical point, there are two distinct phases of supercooled water, a low density liquid (LDL) and a high density liquid (HDL).  
Other scenarios that have been developed to understand supercooled water, and related liquids with tetrahedral structure such as Si and SiO$_2$, are the singularity-free scenario~\cite{Sastry:1996ik,Rebelo:1998hn}
and the critical-point-free (CPF) scenario~\cite{Angell:2008bx}.
The CPF scenario is a variant of the LLCP case, in which the critical point of the LDL-HDL phase transition occurs at such large negative pressure that it has become unobservable due to cavitation of the liquid phase.
See Refs.~\cite{Gallo:2016fd,Handle:2017is,Palmer:2018et} 
for recent overviews of these scenarios.

In 2003, Debenedetti argued that key predictions of the SLC are inconsistent with thermodynamics~\cite{Debenedetti:2003p7278}.  
This critique is based on the observation, illustrated in Fig.~\ref{slc}, that a reentrant liquid spinodal will necessarily intersect the metastable extension of the G-L binodal in the supercooled region.  In the absence of any other phase transitions, the intersection of a liquid spinodal and a G-L binodal must be a critical point.  Furthermore, just like at the high temperature G-L critical point, the liquid spinodal will end at such a spinodal-binodal intersection.  
{\color{black} 
The continuation of the liquid spinodal beyond this intersection thus appears to be precluded on thermodynamic grounds. 
Assuming that the G-L binodal is monotonic in the $T$-$P$ plane, the intersection with the spinodal must occur at a pressure below the triple point, which lies below atmospheric pressure.  In this case, a reentrant spinodal will not be encountered on cooling at ambient and higher $P$, and so cannot explain the observed anomalies of supercooled water.}

In an exchange of comments, Speedy~\cite{Speedy:2004gg} 
and Debenedetti~\cite{debenedetti:2004cm} 
debated this critique of the SLC.  Speedy disagreed with Debenedetti's assertion that a spinodal-binodal intersection must always be a critical point, citing the example of the (non-critical) endpoint expected to occur at negative pressure on the metastable extension of the crystal-liquid binodal where it meets the liquid spinodal in a simple atomic system.  However, in the absence of an example of a non-critical spinodal-binodal intersection involving only fluid phases, 
Debenedetti's critique has been widely accepted.
As a consequence, 
the SLC is now usually excluded from consideration as a viable explanation for the behavior of water and water-like liquids; see e.g. Refs.~\cite{Caupin:2015iw,Gallo:2016fd,Handle:2017is}.

Nonetheless, there has recently been renewed interest in the SLC.
New experiments are making steady progress at quantifying the properties of water at negative pressure, where the predictions of the SLC and other thermodynamic scenarios may be directly tested~\cite{Pallares:2014iu,Caupin:2015iw,Holten:2017ip}.
Also, a recent simulation study of two patchy colloid systems found in both cases a reentrant liquid spinodal at which the liquid vaporizes on cooling at constant $P$~\cite{Rovigatti:2017fb}.  
The analysis of these two systems indicated that each has a closed loop 
G-L coexistence region with a G-L critical point at both high and low $T$. 
{\color{black} In both models, the G-L coexistence curve is not monotonic in the
$T$-$P$ plane, at odds with the monotonicity implicitly assumed in Debenedetti's reasoning~\cite{Debenedetti:2003p7278}.
However,} consistent with Debenedetti's analysis,
the liquid spinodal terminates on cooling at the low-$T$ critical point. 
Recent simulations of the generalized Stillinger-Weber model have also produced evidence of a reentrant liquid spinodal at negative pressure, although it is not yet clear in this case if the spinodal extends to positive $P$ at lower $T$~\cite{Russo:2018jz}.

In addition, it is possible for the CPF scenario to include behavior reminiscent of the SLC.  This was first demonstrated in Ref.~\cite{Poole:1994p4399}, where an ``extended van der Waals" (EVDW) model equation of state for water was shown to generate both the LLCP scenario as well as a scenario that included a SLC-like reentrant spinodal.  This latter case was subsequently associated with the CPF scenario~\cite{Angell:2008bx}.
However, Ref.~\cite{Poole:1994p4399} 
did not present an analysis of all the binodals and spinodals occurring in each scenario, including that with a reentrant spinodal.  
The prospect therefore exists that a more detailed study of the EVDW model may be able to provide examples of spinodal-binodal intersections that can help resolve the thermodynamic viability of the SLC.  

Accordingly, in the present work we re-examine the EVDW model, with the aim of clarifing the behavior of all spinodals and binodals that occur in both the LLCP and CFP scenarios.
As shown below, we find that the EVDW model exhibits several cases where a binodal terminates on a spinodal at a point which is not a conventional critical point.  
In the following, we refer to such points as ``Speedy points", to acknowledge Speedy's recognition of their significance in metastable systems~\cite{Speedy:2004gg}.  
At the same time, we also find that a Speedy point only occurs when there is more than one phase coexistence in the system.
Thus our observations are not in conflict with
Debenedetti's critique of the original SLC, which explicitly considers only the gas-liquid transition~\cite{Debenedetti:2003p7278}.  
As we will see, in the EVDW model it is the presence of the liquid-liquid transition that makes it possible to terminate the gas-liquid binodal in the manner predicted by the SLC.

\begin{figure}
\centerline{\includegraphics[scale=0.20]{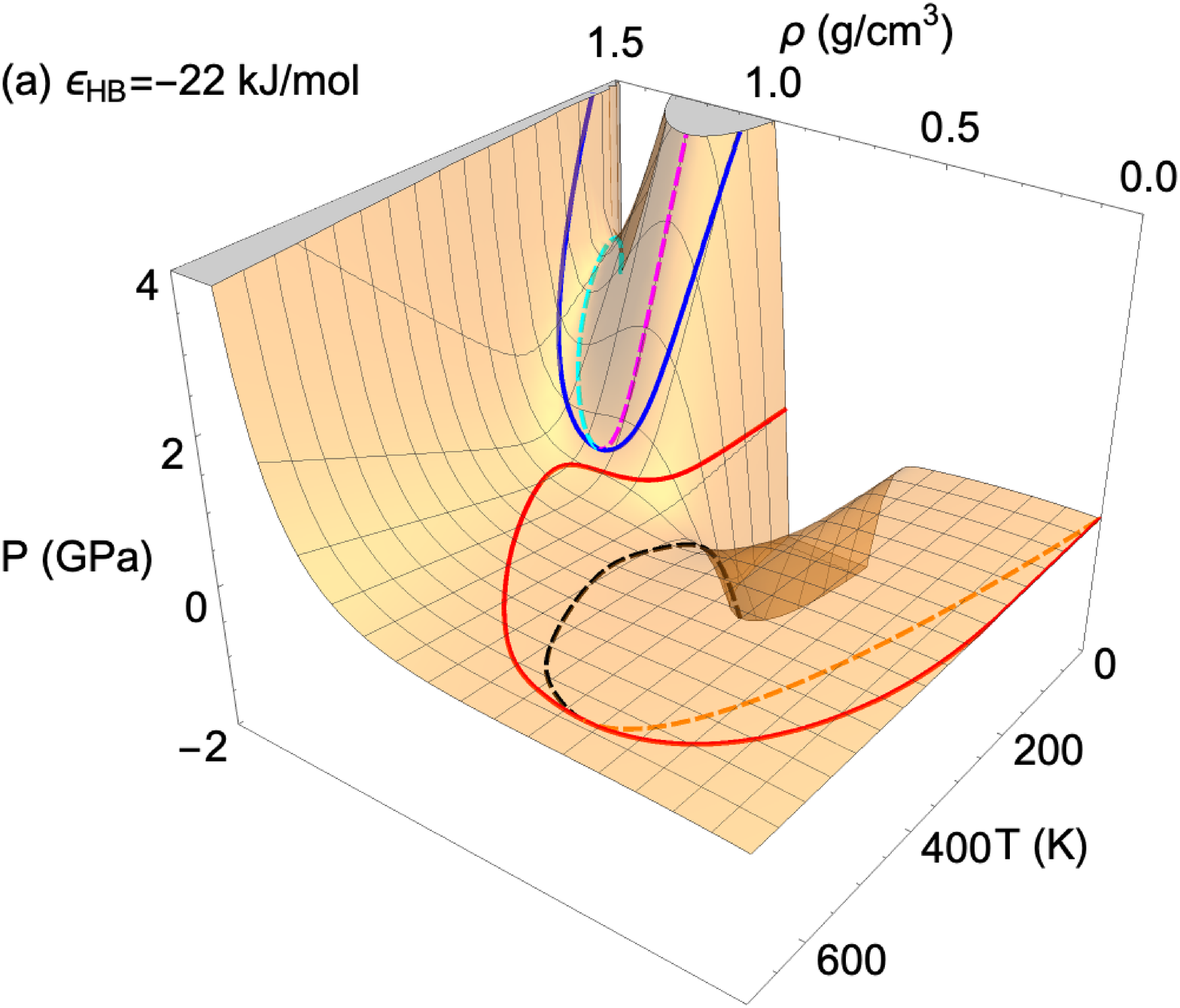}}
\centerline{\includegraphics[scale=0.20]{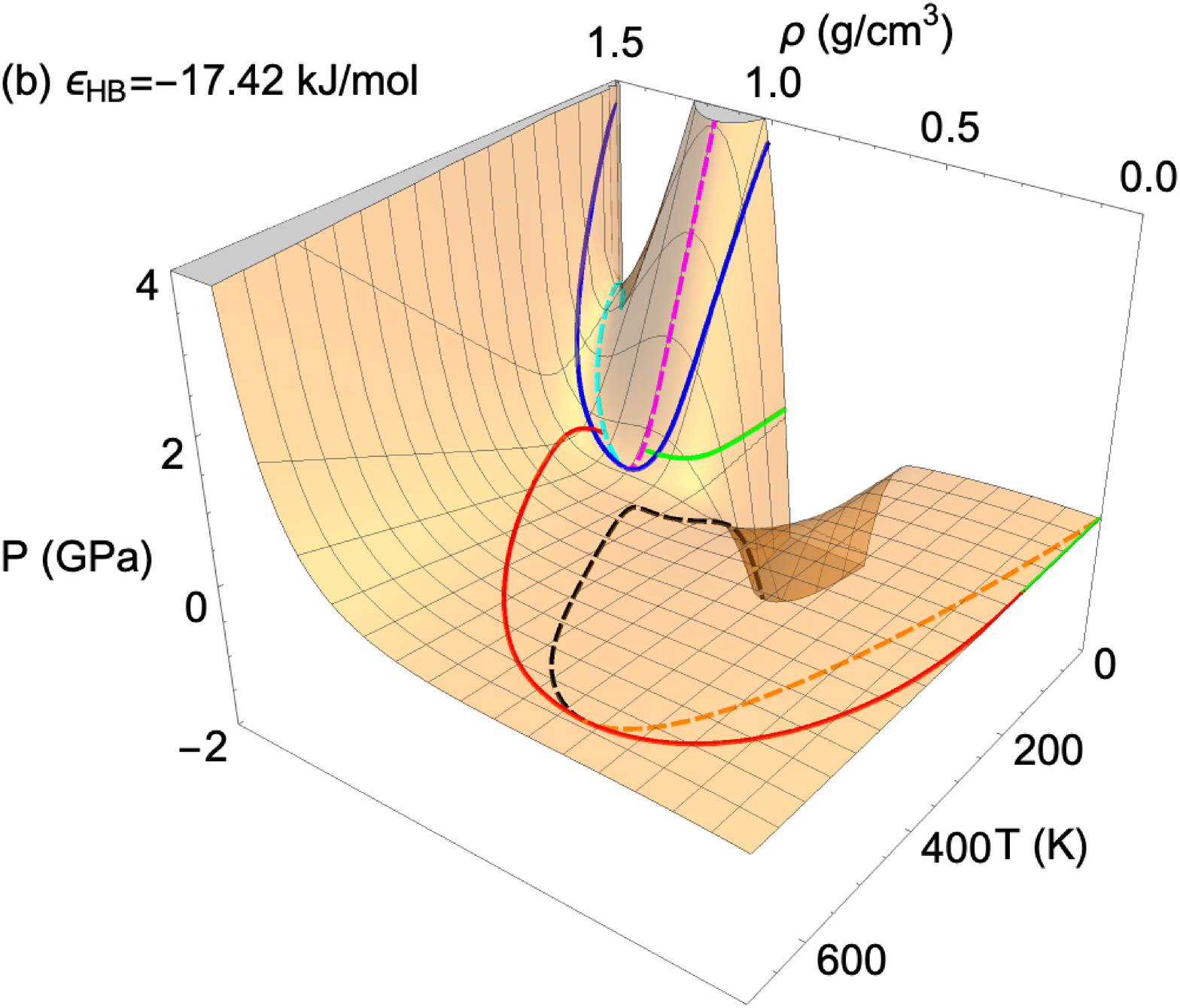}}
\centerline{\includegraphics[scale=0.20]{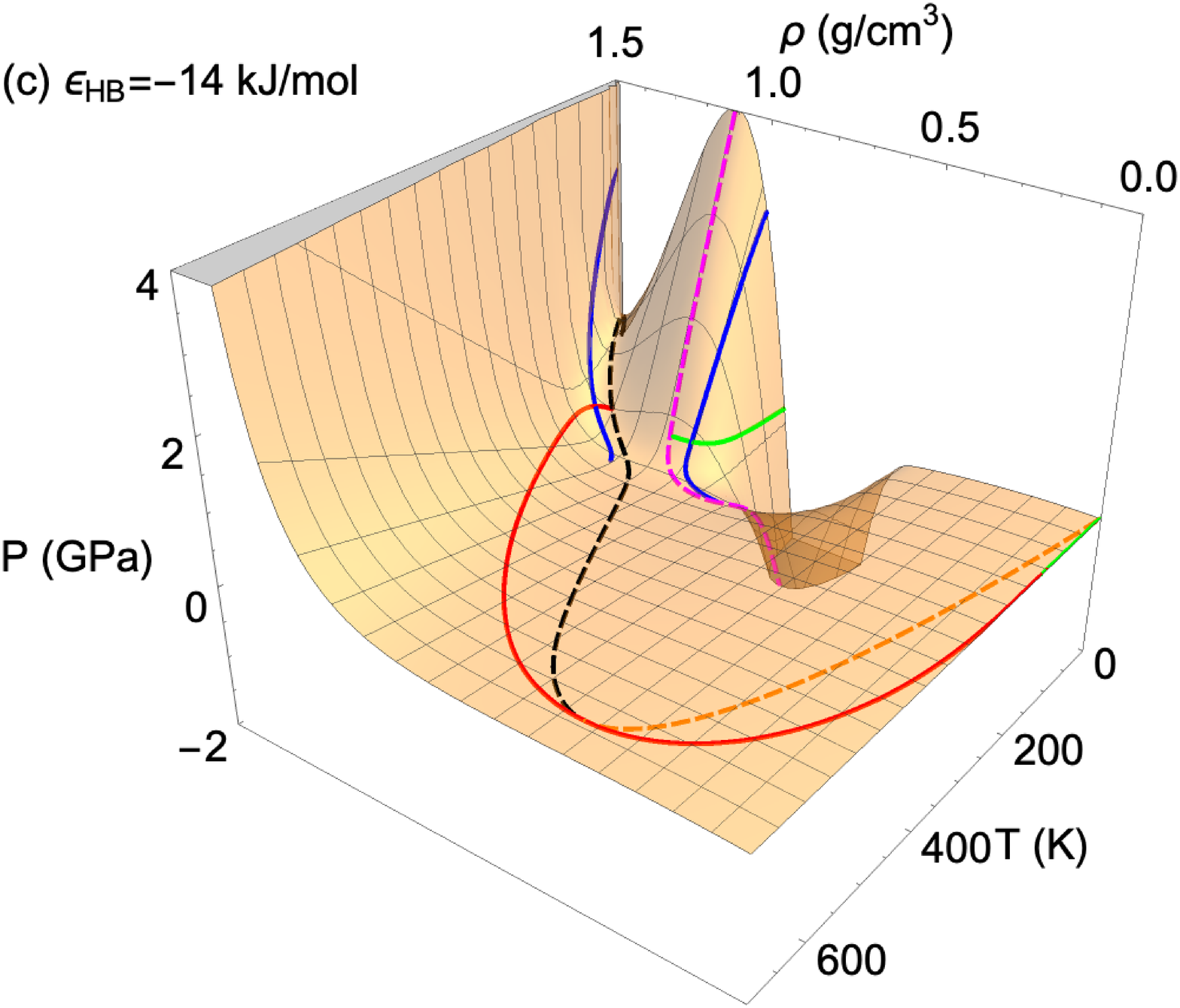}}
\caption{Surface plots of $P(\rho,T)$ for 
(a)~$\epsilon_{\rm HB}=-22$~kJ/mol,
(b)~$\epsilon_{\rm HB}=-17.42$~kJ/mol, and
(c)~$\epsilon_{\rm HB}=-14$~kJ/mol.
Definitions for all lines are given in Table~\ref{table}.}
\label{3d}
\end{figure}

\section{extended van der Waals model}
\label{evdw}

The EVDW model is a simple equation of state that incorporates the physics of both a gas-liquid and liquid-liquid phase transition.  The motivation and details of the EVDW model are fully described in Ref.~\cite{Poole:1994p4399}.  Here we briefly summarize the formulation of the model.
In the EVDW model, the molar Helmholtz free energy $A$ 
is given by, 
\begin{equation}
A(v,T)=A_{\rm VDW}+2A_{\rm HB},
\label{e1}
\end{equation}
where,
\begin{eqnarray}
A_{\rm VDW}&=&-RT\{  \ln[(v-b)/\Lambda^3]+1\}-a^2/v\\
A_{\rm HB}&=&-fRT\ln[\Omega + \exp(-\epsilon_{\rm HB}/RT) ]\\
&&-(1-f)RT\ln(\Omega+1) \nonumber
\end{eqnarray}
and,
\begin{equation}
f=\exp\{-[(v-v_{\rm HB})/\sigma]^2\}.
\label{e2}
\end{equation}
$A_{\rm VDW}$ is the Helmholtz free energy for the van der Waals fluid.  $A_{\rm HB}$ models the influence of hydrogen bonds on the liquid phase.  To describe the free energy contributed by bonds between pairs of neighboring water molecules, each bond is modelled as an independent $(\Omega +1)$-state system.  There are $\Omega$ possible non-hydrogen-bonded states each of zero energy, and one hydrogen bonded state of energy $\ehb<0$.  The factor of 2 in Eq.~\ref{e1} arises because there are two hydrogen bonds per molecule.  
The function $f$ defined in Eq.~\ref{e2} models the fact that only when the molar volume $v$ of the liquid is near the optimal (ice-like) value $v_{\rm HB}$ can all the bonds in the system have the potential to form hydrogen bonds.
In the above equations, $R$ is the gas constant; 
$\Lambda$ is the thermal de Broglie wavelength; 
$v$ is the molar volume in m$^3$mol$^{-1}$;
$a=0.218$~Pa$^{1/2}$m$^3$mol$^{-1}$; 
$b=12.01\times 10^{-6}$~m$^3$mol$^{-1}$; 
$\Omega=\exp(-S_{\rm HB}/R)$ with $S_{\rm HB}=-90$~J\,mol$^{-1}$K$^{-1}$; 
$v_{\rm HB}=19.58\times 10^{-6}$~m$^3$/mol;
and $\sigma=v_{\rm HB}/4$.  
The physical motivation for choosing the values of these parameters is described in Ref.~\cite{Poole:1994p4399}.

To obtain the different thermodynamic scenarios discussed below, we 
vary only the value of $\epsilon_{\rm HB}$.
To evaluate the phase diagram for each case, we first find
the pressure equation of state $P(v,T)$ from $P=-(\partial A/\partial v)_T$.  Spinodals lines are then found as the set of points satisfying $(\partial P/\partial v)_T=0$.  Binodals (i.e. coexistence curves) are obtained from isotherms of $P(v,T)$ using the Maxwell equal-area construction~\cite{Callen}.
In the following, we plot the phase behavior as projected into both the $\rho$-$T$ and $T$-$P$ planes, 
where the mass density $\rho=M/v$ and $M=18.016$~g/mol is the molar mass of water.

{\color{black} We note that since $P$ is related to the
derivative of $A$ with respect to $v$, the contribution of $A_{HB}$ to the $v$ dependence of $P$, through the gaussian function $f$, has the form of the first Hermite polynomial (which also describes the shape of the wave function of the first exited state of the one-dimensional quantum harmonic oscillator).  As we will see below, the effect of this form is to enhance the thermodynamic stability of the system when $v$ is close to $v_{\rm HB}$.  At low $T$, this term adds an additional ``van der Waals loop" to the pressure equation of state, creating a liquid-liquid transition distinct from the gas-liquid transition.}

\begin{table*}
  \begin{tabular}{ | c | c | c |} \hline
    abbreviation & line color \& type & definition  \\ \hline \hline
    \spin(G) & orange dashed & gas spinodal originating in the G\hh L critical point \\ \hline
    \spin(L) & black dashed & liquid spinodal originating in the G\hh L critical point \\ \hline
    \spin(LDL) & magenta dashed & LDL spinodal originating in a liquid-liquid critical point,\\ 
    & & or that forms the boundary of an isolated region of LDL stability \\ \hline
    \spin(HDL) & cyan dashed & HDL spinodal originating in a LDL\hh HDL critical point\\ \hline
    \coex(G\hh L) & red solid & G\hh L binodal originating in the G\hh L critical point \\ \hline
    \coex(LDL\hh HDL) & blue solid & LDL\hh HDL binodal \\ \hline
    \coex(G\hh LDL) & green solid & G\hh LDL binodal that is distinct from \coex(G\hh L)\\ \hline
    TMD & black dotted & temperature of maximum density line \\ \hline
  \end{tabular}
  \caption{Definitions and abbreviations for all phase diagram features presented in this work.  In all phase diagram plots, the indicated colors and line types are used to plot the corresponding features.}
  \label{table}
  \end{table*}

\section{Phase diagrams}

We study the variation in the properties of the EVDW model over the range $\ehb=-22$ to $-14$~kJ/mol.  The changes in the equation of state surface $P(\rho,T)$ are shown in Fig.~\ref{3d}.
The variation in the $\rho$-$T$ and $T$-$P$ phase diagrams is shown in 
Figs.~\ref{e22}\hh \ref{e14}.  
In all our plots, binodals are plotted as solid lines and spinodals are plotted as dashed lines.  The TMD is plotted as a thin dotted line.  To identify specific binodals and spinodals, we use the following notation:  
\coex($x$\hh $y$) refers to the binodal representing the coexistence of phases $x$ and $y$,  
and \spin($x$) refers to the spinodal representing the limit of metastability of phase $x$.  

{\color{black}  
As $T\to 0$, there are always three distinct phases in our system:  gas (G), LDL, and HDL.
{\color{black}  The EVDW model does not consider crystal phases, and so for a real system the behavior discussed here will usually correspond at low $T$ to states that are metastable relative to a crystalline phase.
Also,} as this is a mean-field model, our formulation does not
discriminate between ergodic and non-ergodic states.  Hence our results report the equilibrium properties of the system even as $T \to 0$, where any realistic system would be below its glass transition temperature.}
When $T$ is sufficiently large so that coexistence is not possible between distinct LDL and HDL phases, we simply refer to the liquid phase (L).  
Depending on the value of $\ehb$, we find that up to four distinct spinodals and three distinct binodals are observed.
A complete list of all the spinodals and binodals identified and plotted in Figs.~\ref{3d}\hh \ref{e14} is given in Table~\ref{table}, along with the color and line type used to plot each.

As pointed out in Ref.~\cite{Poole:1994p4399}, 
there are two regimes of phase behavior, depending on whether $\ehb$ is less or greater than $\ehb^0=-16.55$~kJ/mol.  
The regime  $\ehb<\ehb^0$ 
is illustrated in Figs.~\ref{e22}\hh \ref{e166} and corresponds to the LLCP scenario.  
In this case there are distinct critical points terminating separate \coex(G\hh L) and \coex(LDL\hh HDL) binodals, and two separate regions of instability occur in the $\rho$-$T$ plane.
LDL is found between these two regions of instability at low $T$, but there is always a continuous path leading out of the LDL phase to the high-$T$ liquid phase L.
Fig.~\ref{e22} presents the case where the LDL\hh HDL critical point occurs at positive pressure.
If the pressure of the LDL\hh HDL critical point drops below \coex(G\hh L), as shown in Fig.~\ref{e17}, then a triple point involving the phases G, LDL, and HDL occurs, and a new binodal \coex(G\hh LDL) appears that is distinct from \coex(G\hh L).
We also note that in the narrow range $-17.42$~kJ/mol~$< \ehb<\ehb^0$ we observe {\it two} liquid-liquid critical points; see Fig.~\ref{e166}.  We discuss this unexpected phenomenon in Section~\ref{other}.

The regime $\ehb>\ehb^0$ 
is illustrated in Figs.~\ref{e165} and \ref{e14} and corresponds to the CPF scenario.  
In this case, 
the two separate unstable regions in the $\rho$-$T$ plane merge into a single contiguous unstable region that completely encloses an isolated ``pocket" of stability associated with the LDL phase.  In this case, there is no thermodynamic path from the LDL phase to another phase that does not involve a first-order phase transition.  The only critical point that occurs in this regime is the G\hh L critical point.

\section{Intersection of a spinodal and a binodal}
\label{speedy}

\begin{figure}
\centerline{\includegraphics[scale=\xx]{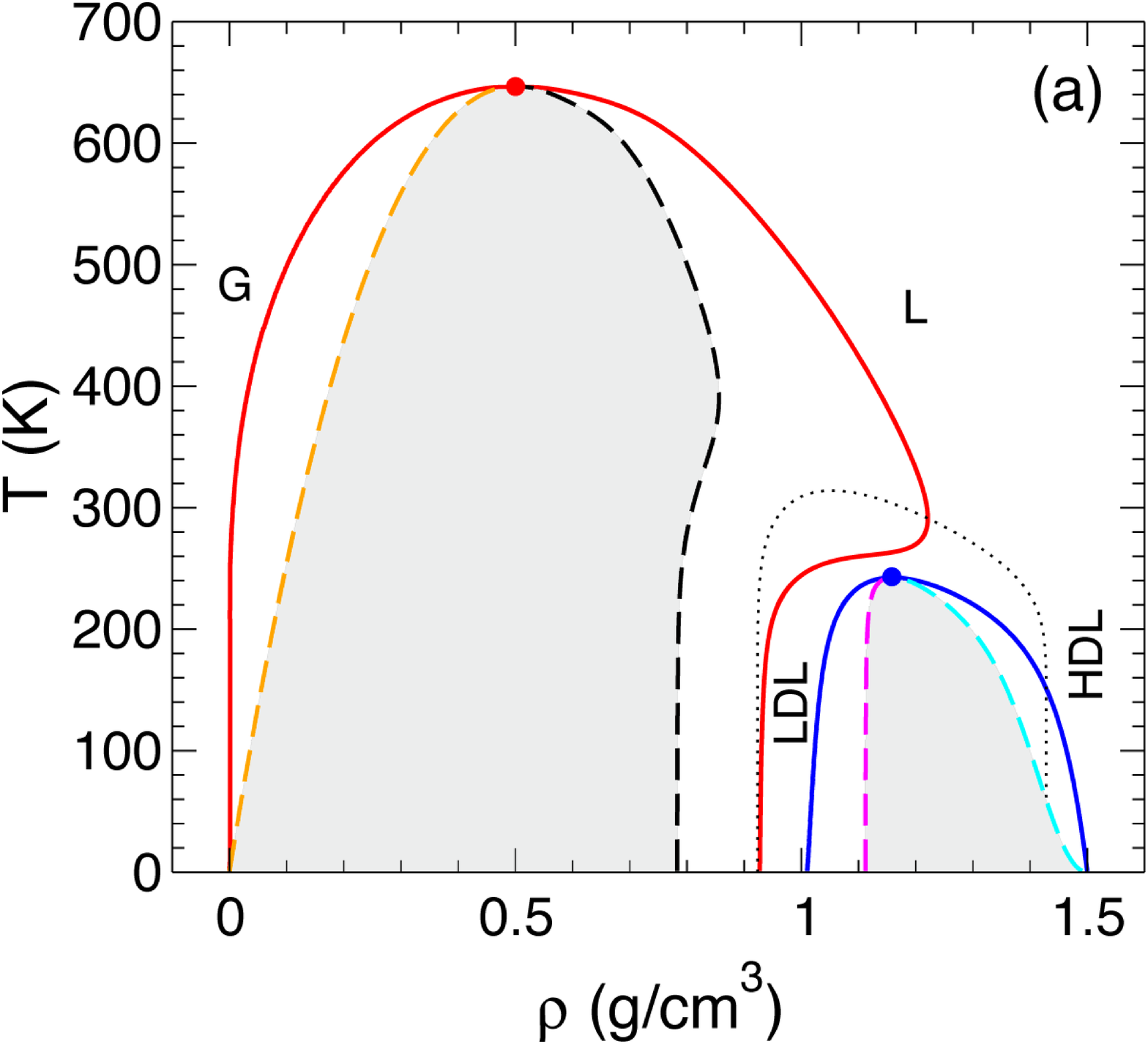}}
\centerline{\includegraphics[scale=\x]{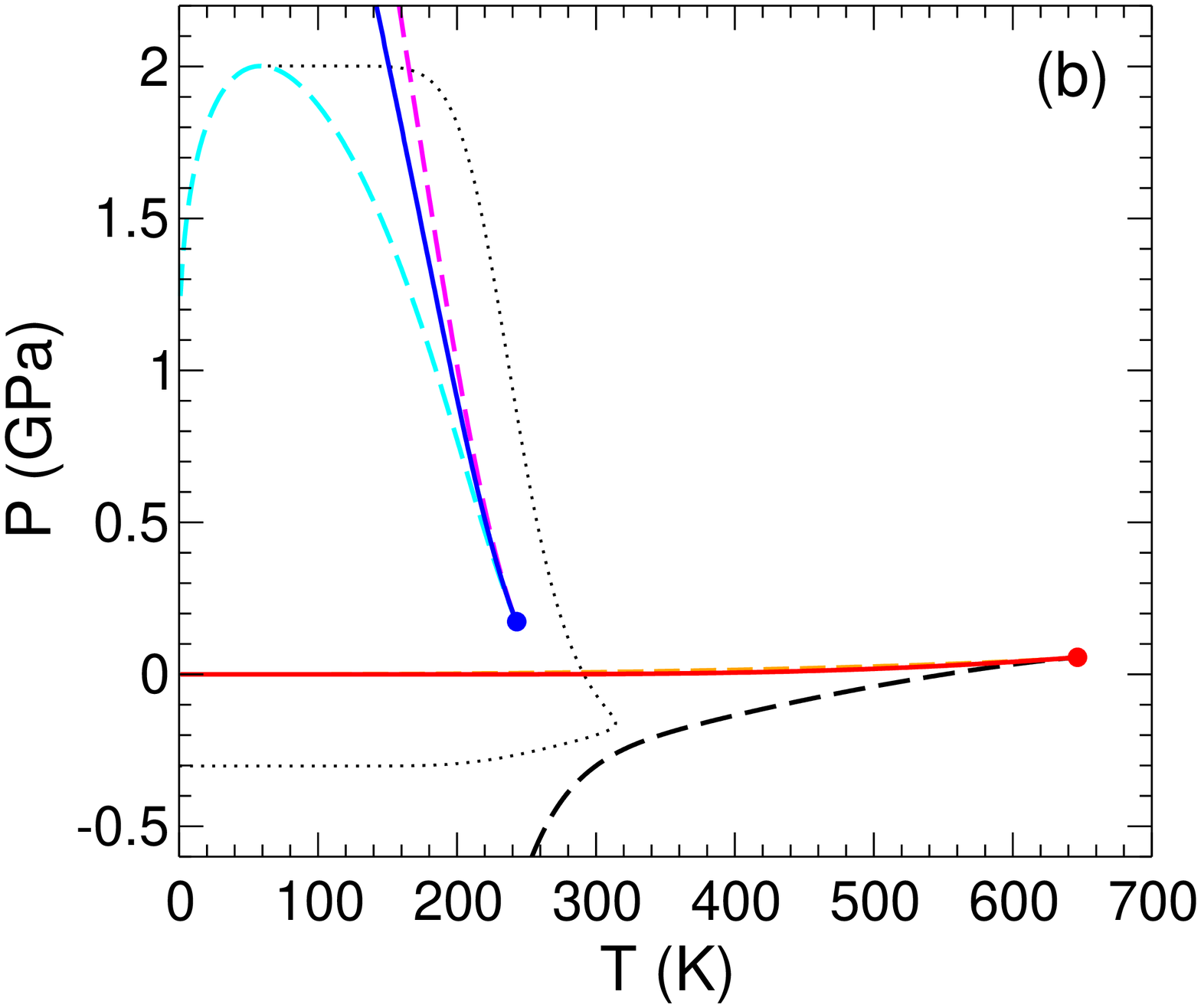}}
\caption{(a) $\rho$-$T$ and (b) $T$-$P$ phase behavior 
for $\epsilon_{\rm HB}=-22$~kJ/mol, for which 
the model predicts a 
LLCP at positive pressure. 
Definitions for all lines are given in Table~\ref{table}. Filled circles are critical points. 
Unstable regions are shaded grey in (a).}
\label{e22}
\end{figure}

\begin{figure}
\centerline{\includegraphics[scale=\xx]{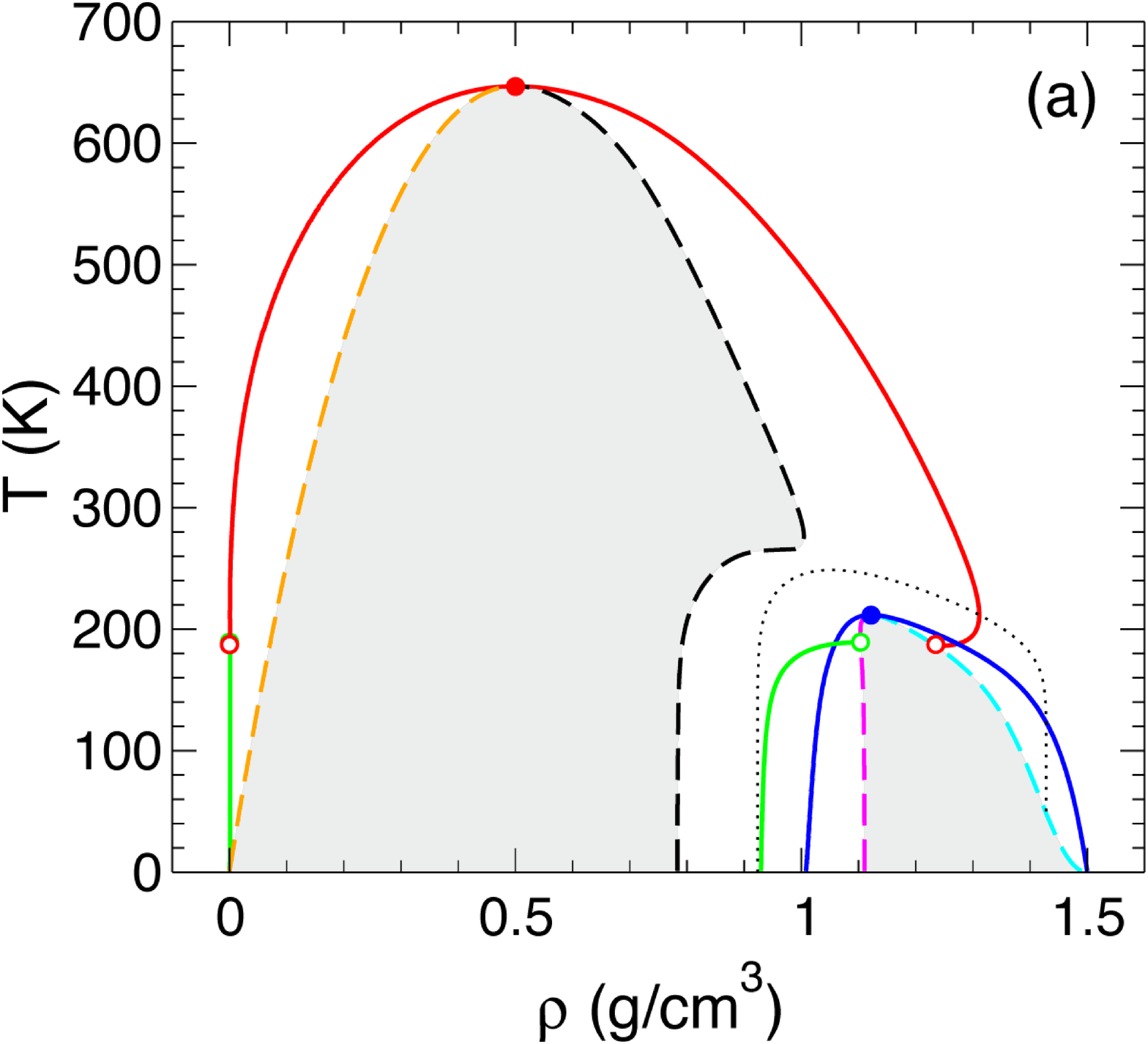}}
\centerline{\includegraphics[scale=\x]{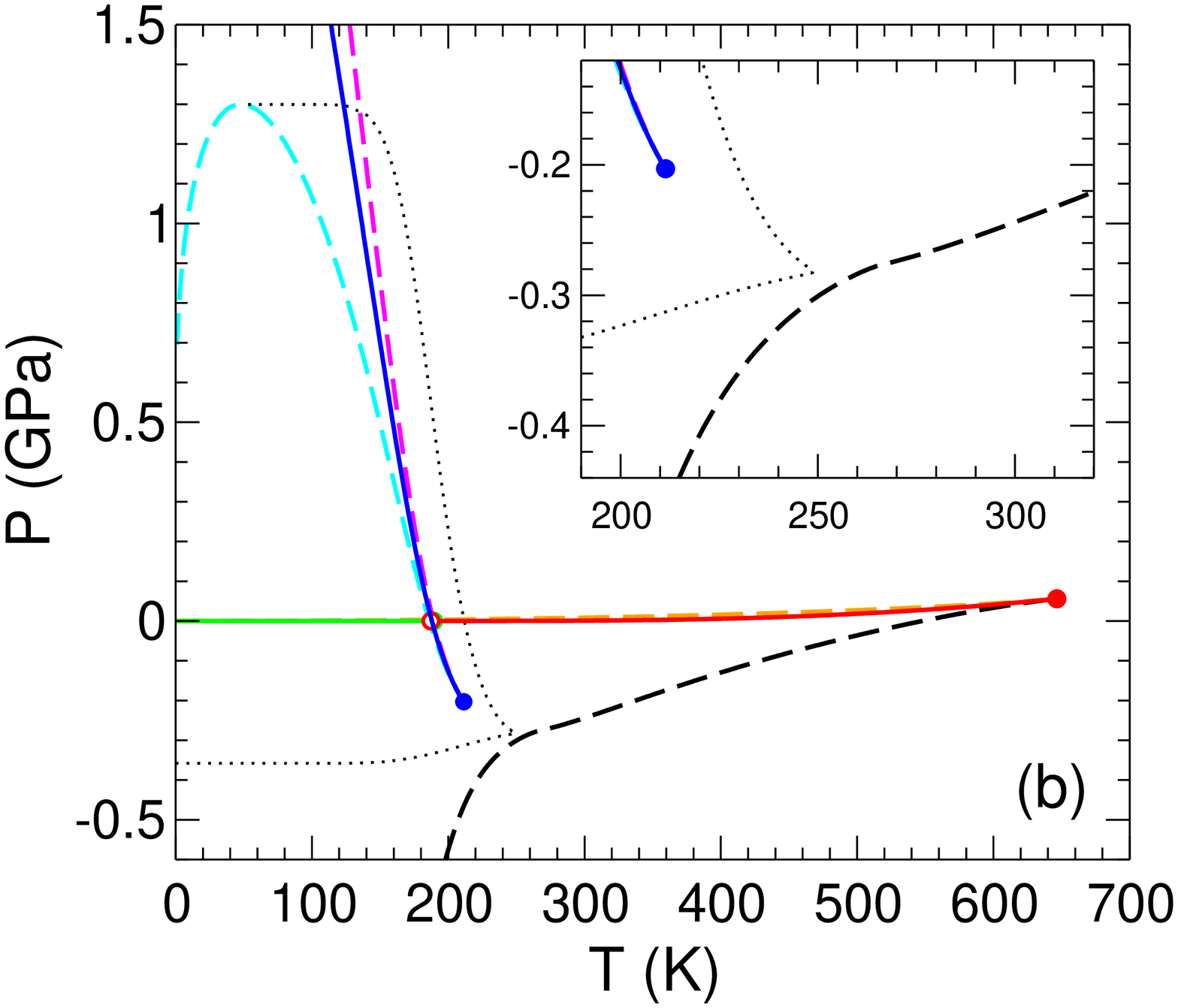}}
\caption{(a) $\rho$-$T$ and (b) $T$-$P$ phase behavior 
for $\epsilon_{\rm HB}=-17.42$~kJ/mol, for which
the model predicts a 
LLCP at negative pressure.
Definitions for all lines are given in Table~\ref{table}. Filled circles are critical points and open circles are Speedy points. Unstable regions are shaded grey in (a).}
\label{e17}
\end{figure}

\begin{figure}
\centerline{\includegraphics[scale=\xx]{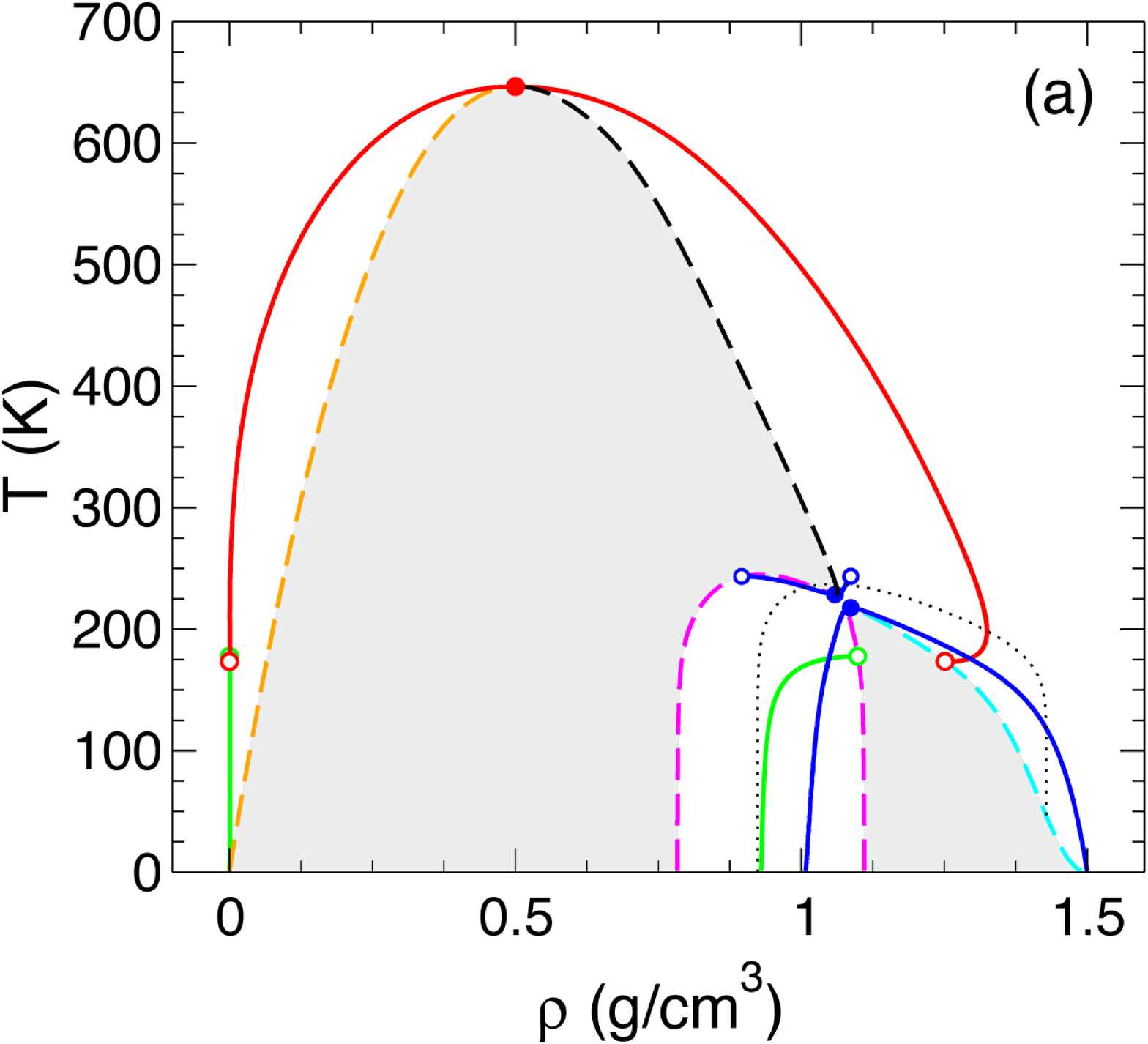}}
\centerline{\includegraphics[scale=\x]{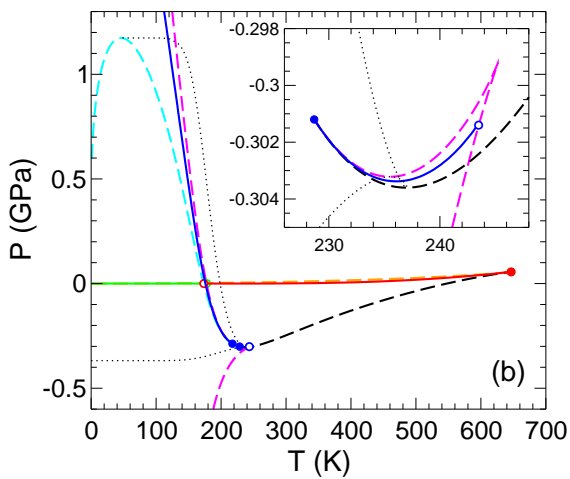}}
\caption{(a) $\rho$-$T$ and (b) $T$-$P$ phase behavior 
for $\epsilon_{\rm HB}=-16.6$~kJ/mol, for which  
the model predicts two
liquid-liquid critical points at negative pressure.
Definitions for all lines are given in Table~\ref{table}. Filled circles are critical points and open circles are Speedy points. Unstable regions are shaded grey in (a).}
\label{e166}
\end{figure}

\begin{figure}
\centerline{\includegraphics[scale=\xx]{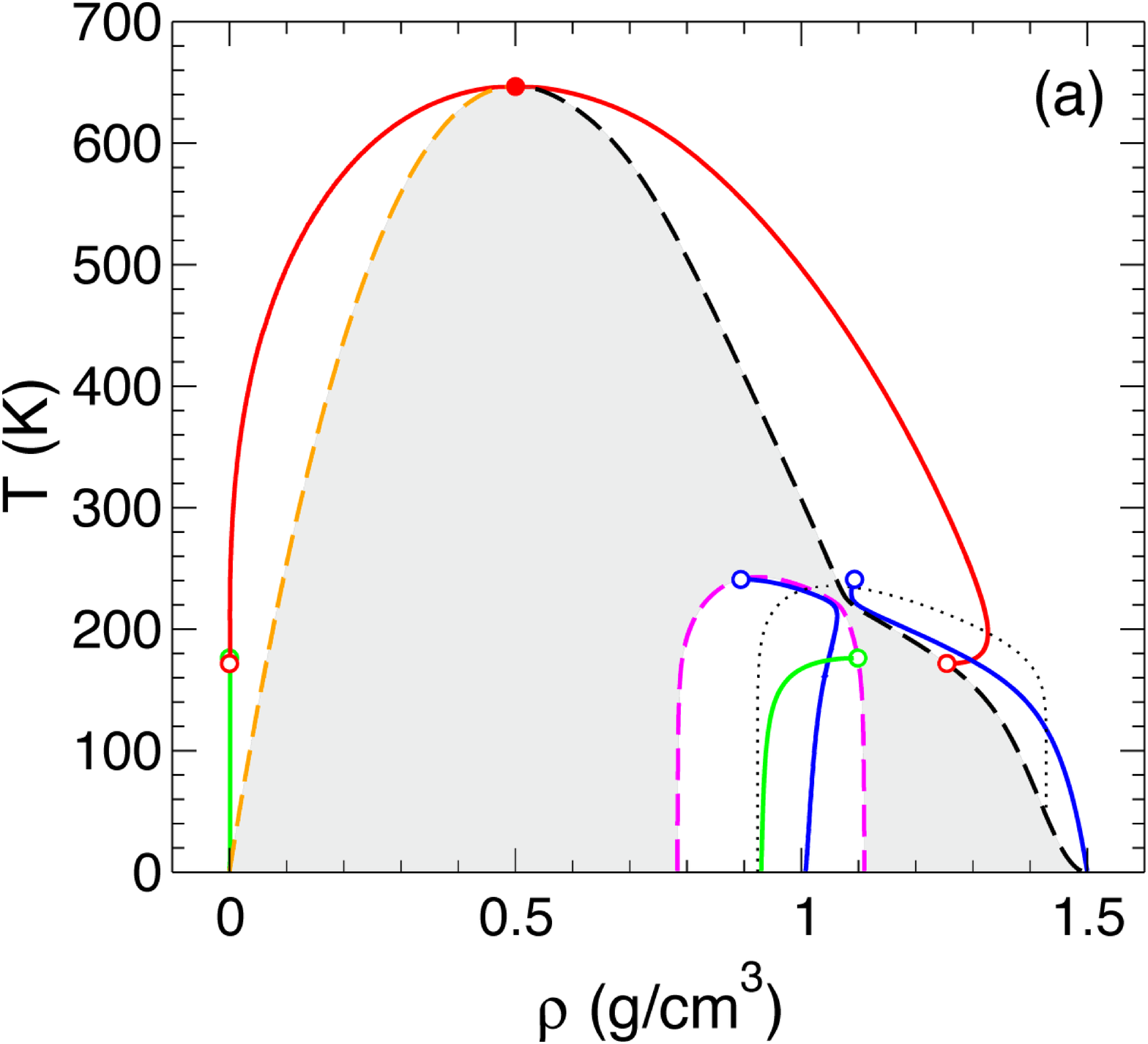}}
\centerline{\includegraphics[scale=\x]{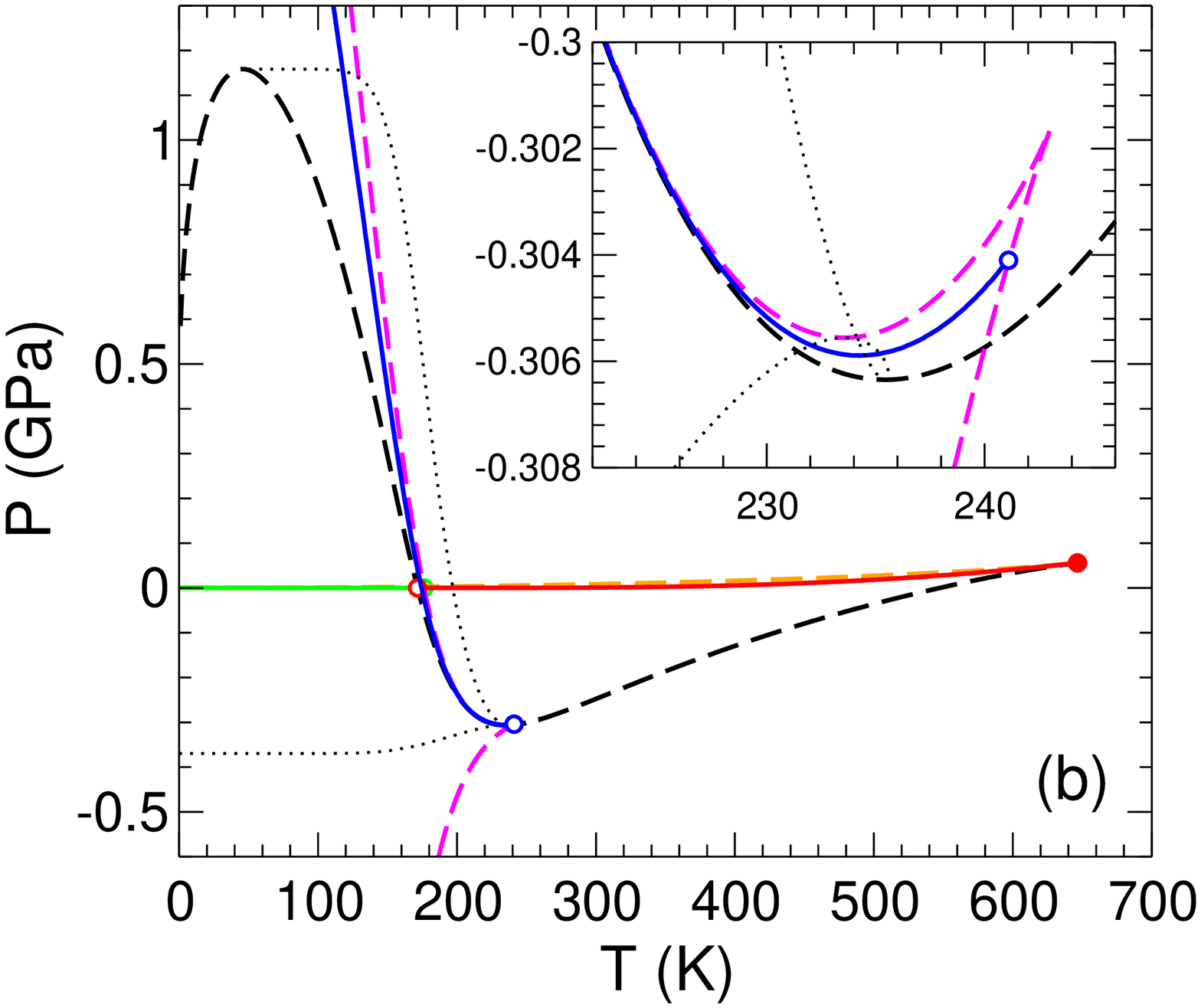}}
\caption{(a) $\rho$-$T$ and (b) $T$-$P$ phase behavior 
for $\epsilon_{\rm HB}=-16.5$~kJ/mol, for which  
the model predicts a 
CPF scenario.
Definitions for all lines are given in Table~\ref{table}. Filled circles are critical points and open circles are Speedy points. Unstable regions are shaded grey in (a).}
\label{e165}
\end{figure}

\begin{figure}
\centerline{\includegraphics[scale=\xx]{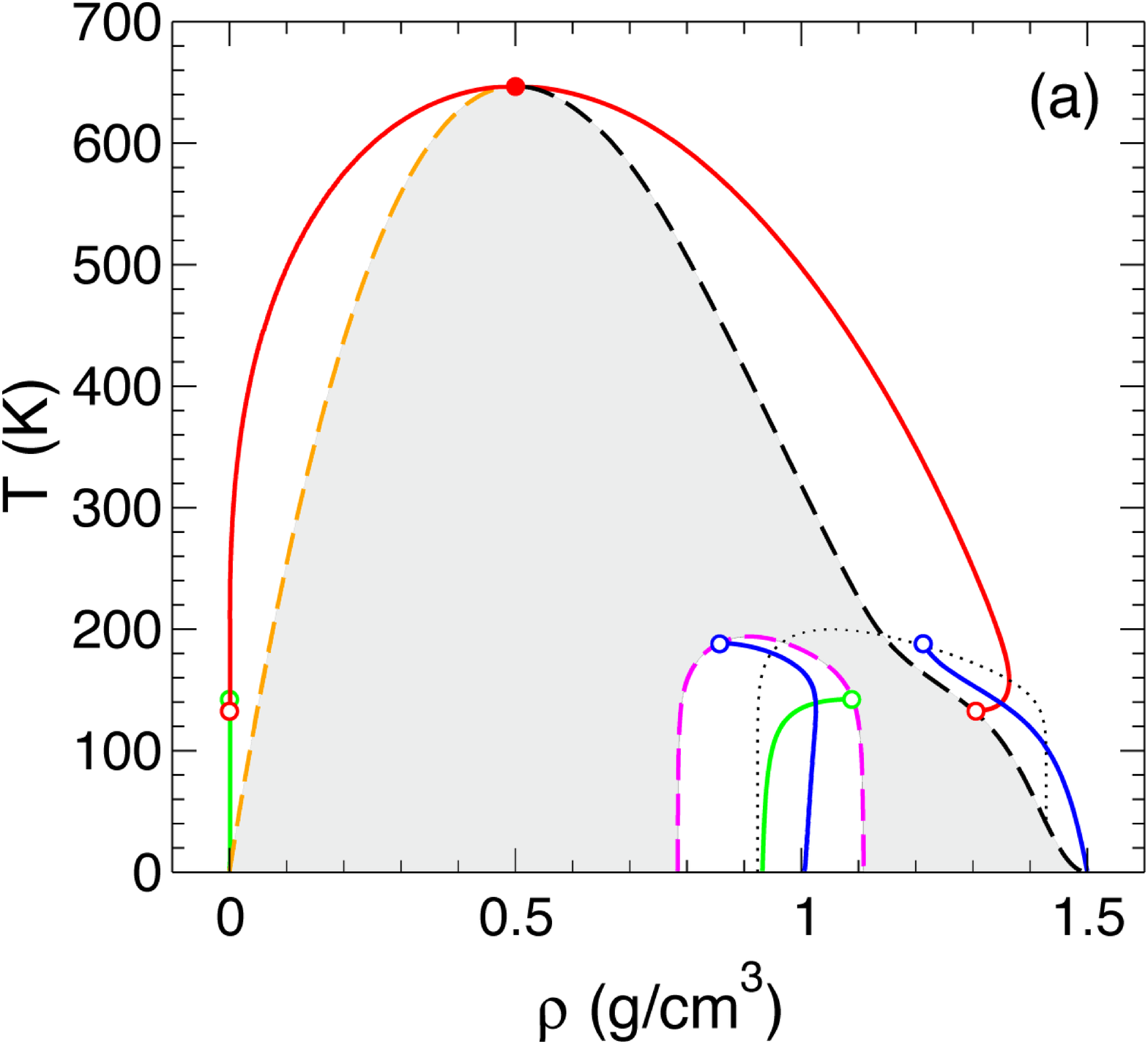}}
\centerline{\includegraphics[scale=\x]{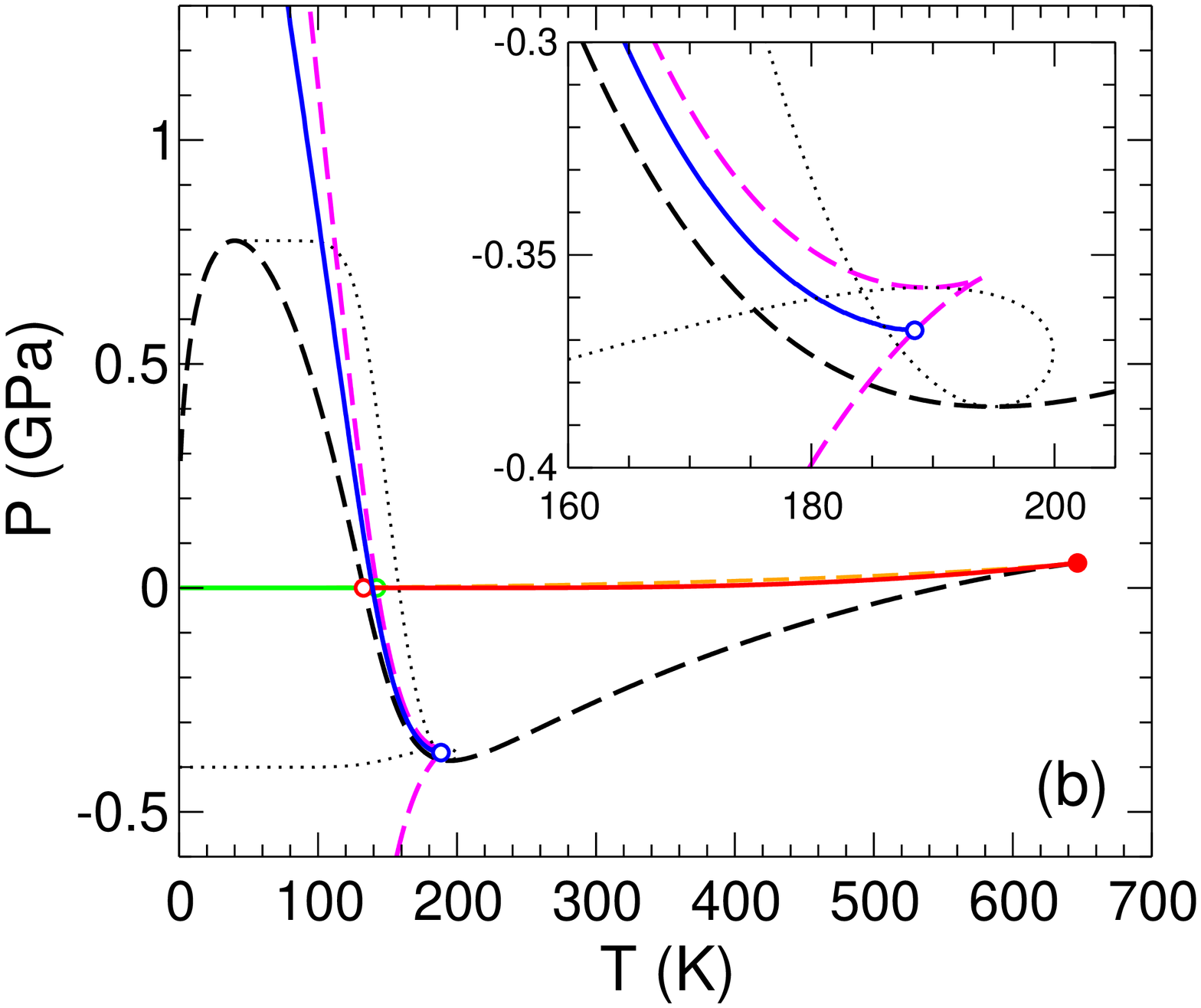}}
\caption{(a) $\rho$-$T$ and (b) $T$-$P$ phase behavior 
for $\epsilon_{\rm HB}=-14$~kJ/mol, for which
the model predicts a 
CPF scenario.
Definitions for all lines are given in Table~\ref{table}. 
Filled circles are critical points and open circles are Speedy points. Unstable regions are shaded grey in (a).}
\label{e14}
\end{figure}

Figs.~\ref{e22}\hh \ref{e14} illustrate several instances of Speedy points, where a binodal ends on a spinodal at a point that is not a critical point.
For example, when $\ehb=-17.42$~kJ/mol (see Fig.~\ref{e17}) the \coex(G\hh L) 
binodal ends when its high density L branch touches \spin(HDL).  
The L phase that coexists with G under these conditions becomes unstable at \spin(HDL) and so \coex(G\hh L) is not defined at lower $T$, even though the G phase remains well-defined and stable.  In the low-$T$ limit, it is the LDL phase that coexists with G, as defined by a distinct \coex(G\hh LDL) binodal.  In a similar manner, the \coex(G\hh LDL) binodal ends as $T$ increases when its LDL branch encounters \spin(LDL).  

\begin{figure}
\centerline{\includegraphics[scale=\x]{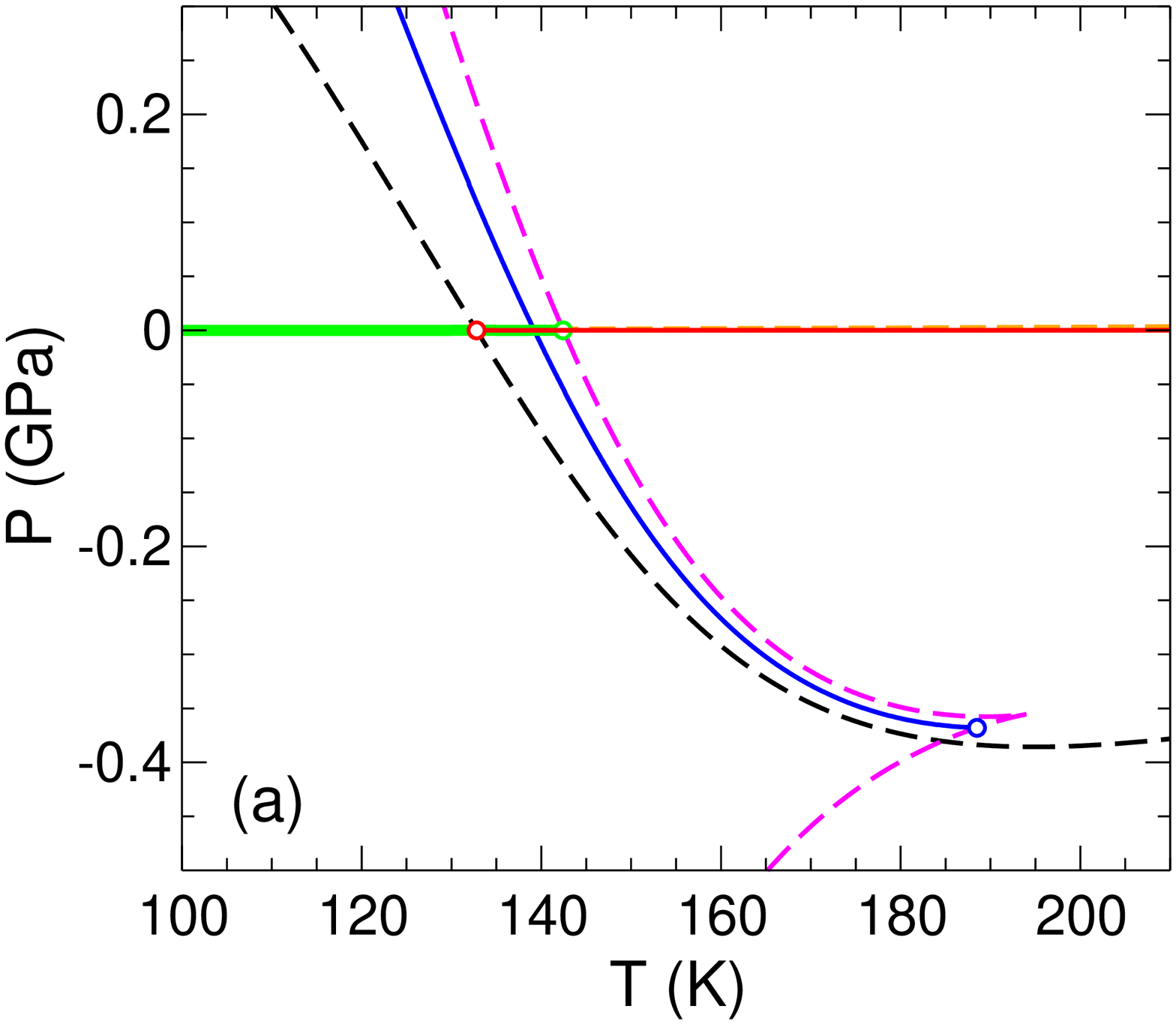}}
\centerline{\includegraphics[scale=0.25]{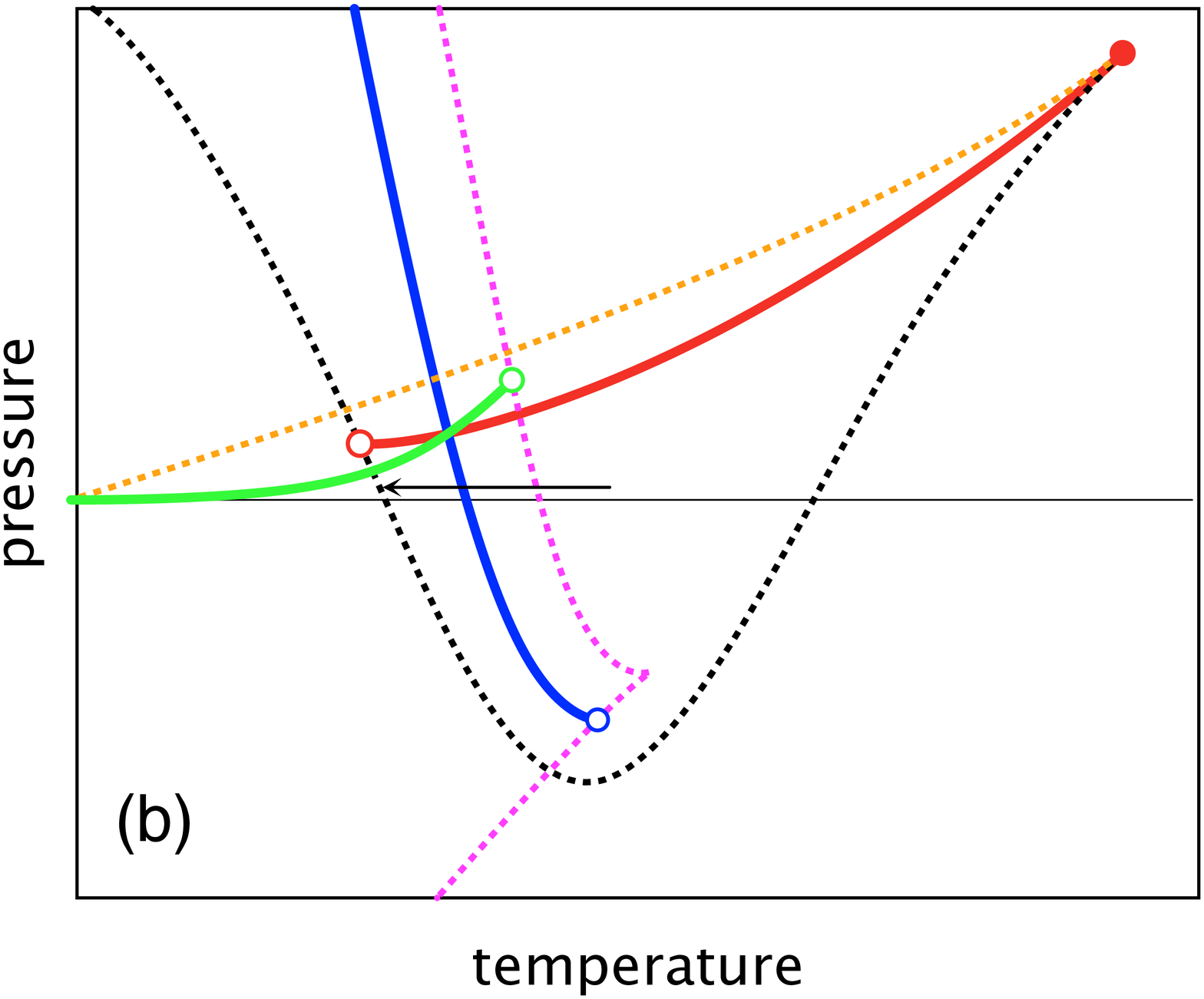}}
\caption{(a) Close-up of phase behavior for 
$\epsilon_{\rm HB}=-14$~kJ/mol.  
(b) Schematic of CPF scenario corresponding to $\epsilon_{\rm HB}=-14$~kJ/mol.  
For both (a) and (b), the definitions for all lines are given in Table~\ref{table}. 
Filled circles are critical points and open circles are Speedy points.  {\color{black}The horizontal arrow identifies an isobaric cooling path for the liquid where the most stable phase of the system beyond the spinodal \spin(L) is the gas phase.  That is, cavitation of the liquid on cooling at positive pressure is thermodynamically permitted along this path.}}
\label{cpf}
\end{figure}

When $\ehb=-14$~kJ/mol 
(see Figs.~\ref{e14} 
and \ref{cpf}), all three binodals end at Speedy points.  In the $T$-$P$ plane, \coex(G\hh L) ends as $T$ decreases on \spin(L), 
and both \coex(G\hh LDL) and \coex(LDL\hh HDL) end at different points on \spin(LDL).  
Fig.~\ref{cpf}(a) 
shows a close-up of the region of the $T$-$P$ plane in which these three Speedy points occur.  
In Fig.~\ref{cpf}(b) we present a schematic version of the same behavior to clarify the interrelationship of the spinodals and binodals in the vicinity of the triple point, since these relationships are difficult to see on the scale of Fig.~\ref{cpf}(a).

The endpoint of \coex(LDL\hh HDL) on \spin(LDL) is particularly clear because the shape of both branches of the \coex(LDL\hh HDL) 
binodal in the $\rho$-$T$ plane, and their relationship to nearby spinodals, is readily visualized.  
As shown in Fig.~\ref{e14}(a), 
the LDL branch of \coex(LDL\hh HDL) 
ends on \spin(LDL), beyond which the LDL phase is unstable.  The HDL branch of \coex(LDL\hh HDL) thus also ends, since there is no LDL phase for HDL to coexist with under these conditions.  However, the HDL phase at this endpoint is a well-defined and stable thermodynamic state.   
The Speedy point that terminates the
\coex(LDL\hh HDL) binodal is therefore a point at which one of the two phases involved in the coexistence exhibits diverging fluctuations and becomes unstable at its spinodal, while the other phase remains stable and is not near its spinodal.  This behavior is in contrast to a conventional critical point, at which both coexisting phases simultaneously exhibit diverging fluctuations as they reach the end of the binodal, where the spinodals for each phase also converge.
As we see in Fig.~\ref{e14}(a), the signature in the $\rho$-$T$ plane of the Speedy point for \coex(LDL\hh HDL) is an ``open binodal".  That is, the binodal does not enclose a well-defined region of the $\rho$-$T$ plane, as it would if it were terminated by a critical point.

The termination of \coex(G\hh L) on \spin(L) 
shown in Figs.~\ref{e14} and \ref{cpf}
is the relevant case for clarifying the thermodynamic viability of the SLC.  When $\ehb=-14$~kJ/mol, \spin(L) forms a continuous metastability limit for the liquid phase, starting from the G\hh L critical point and extending all the way to $T=0$.  The intersection in the $T$-$P$ plane of \coex(G\hh L) and \spin(L) is an example of a gas-liquid binodal which meets the liquid spinodal emanating from the gas-liquid critical point.  This is exactly the intersection which is predicted by the SLC and which is critiqued in Ref.~\cite{Debenedetti:2003p7278}.  As we see in Fig.~\ref{e14}, this intersection is not a critical point.  The G-L binodal remains open at this point, and although the liquid phase becomes unstable, the gas phase remains a distinct and well-defined phase.  Furthermore, \spin(L) is a reentrant spinodal in the $T$-$P$ plane, returning from negative to positive $P$ as $T$ decreases, precisely as proposed in the SLC.  The behavior displayed in Figs.~\ref{e14} and \ref{cpf} thus demonstrates that the predictions of the SLC are consistent with thermodynamics and may be realized in a system that exhibits both a gas-liquid and liquid-liquid phase transition.  

\begin{figure}
\centerline{\includegraphics[scale=\xx]{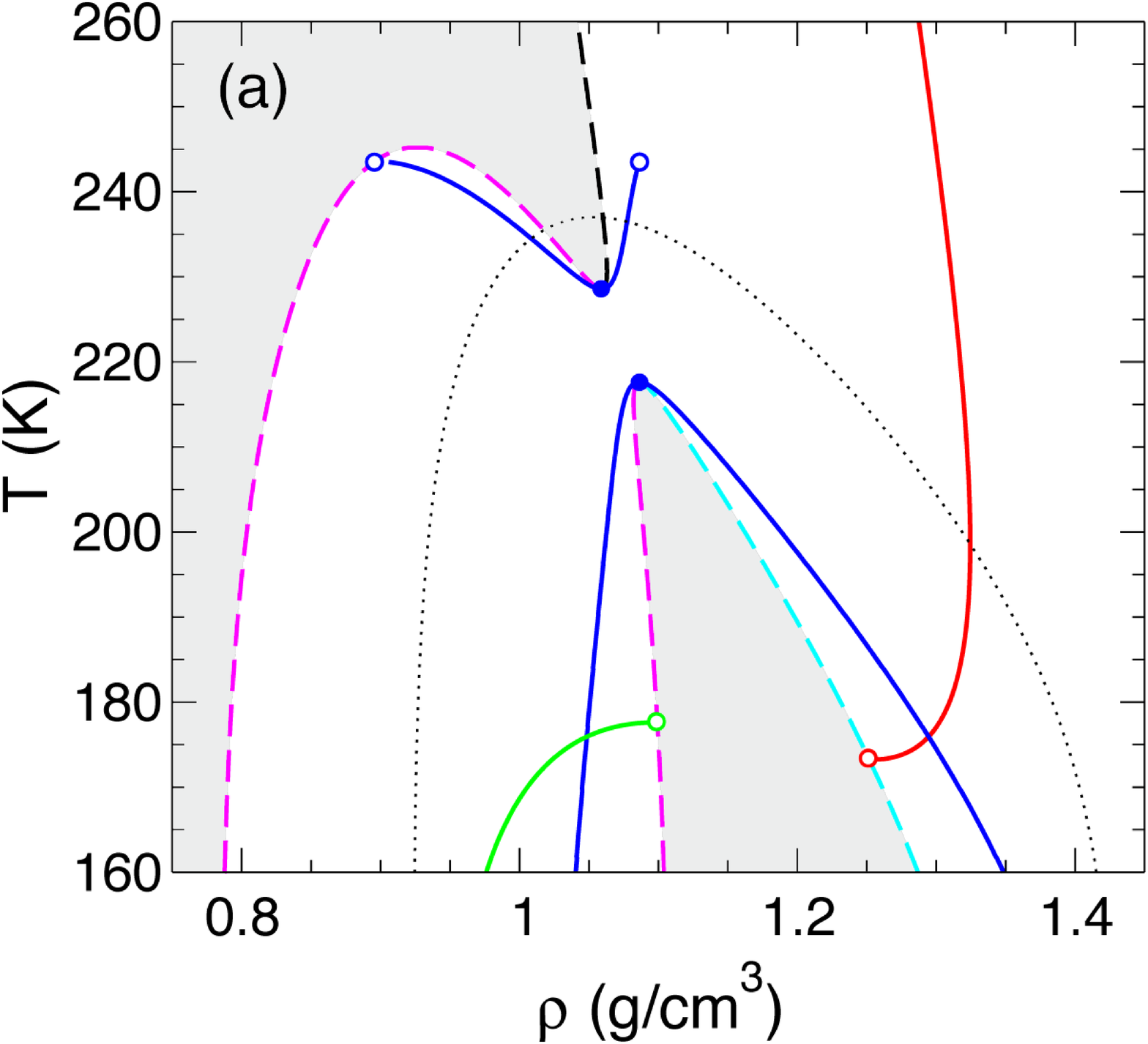}}
\centerline{\includegraphics[scale=\xx]{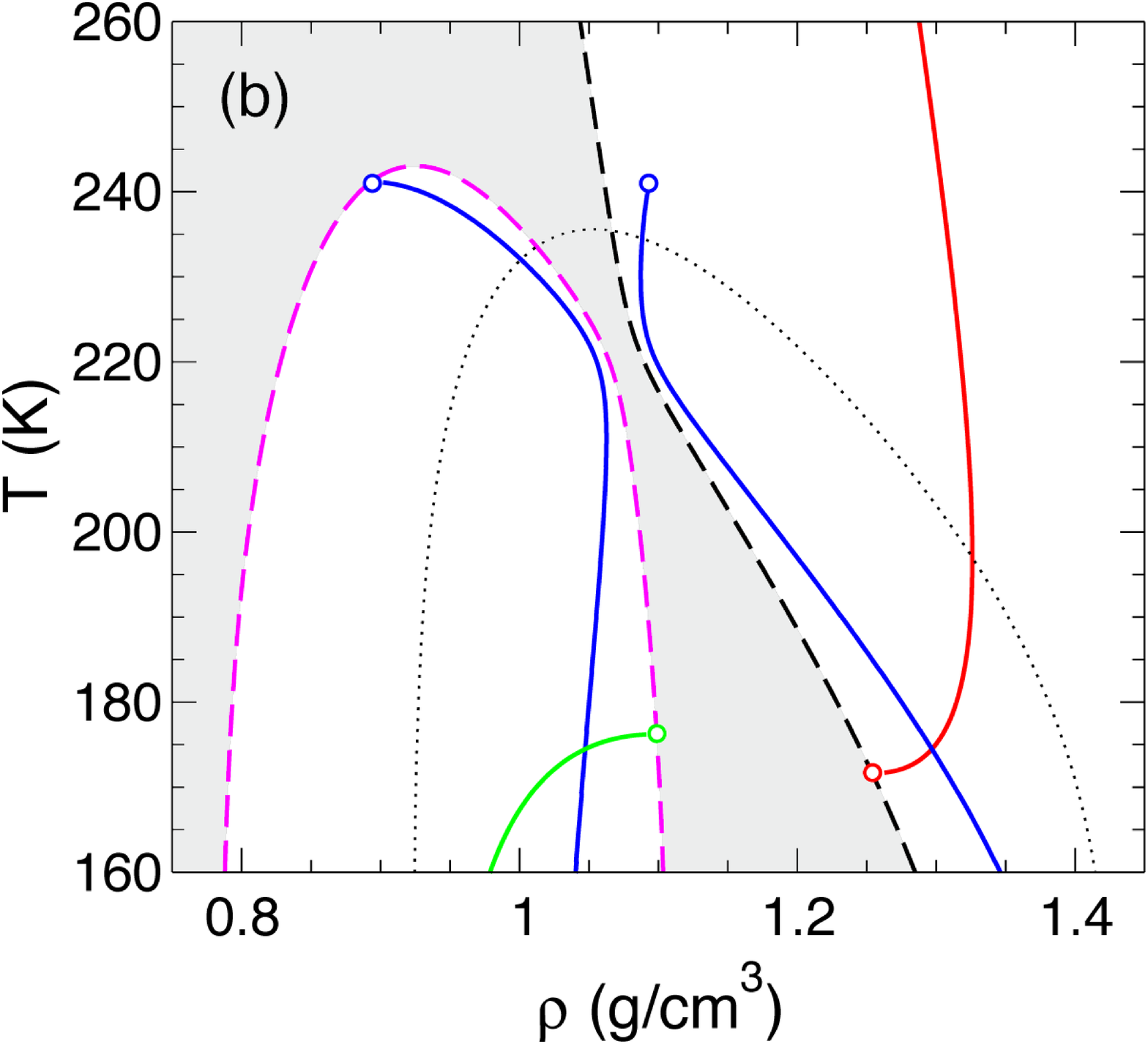}}
\caption{Close-up of phase behavior for 
(a) $\epsilon_{\rm HB}=-16.6$~kJ/mol and (b) $\epsilon_{\rm HB}=-16.5$~kJ/mol.
For both (a) and (b), the definitions for all lines are given in Table~\ref{table}. 
Filled circles are critical points and open circles are Speedy points. Unstable regions are shaded grey.}
\label{newcrit}
\end{figure}

Our results emphasize that a spinodal is defined only as a metastability limit of a single phase.
Information about the new phase that is created beyond the spinodal instability is not encoded in the spinodal itself.  
For example, consider the $T=400$~K isotherm in Fig.~\ref{e14}(a).  
In a system of fixed volume, a single homogeneous L phase can exist for any $\rho$ greater than that of \spin(L).  If $\rho$ decreases to values less than that of \spin(L), then a fixed-volume system at $400$~K must decay into a system with coexisting L and G phases.  However, at $T=50$~K, a fixed-volume system will decay into a system with coexisting HDL and LDL phases when $\rho$ decreases through the density of \spin(L).  That is, even though \spin(L) is a single continuous spinodal, moving to a state point lying beyond \spin(L) produces a different result, depending on $T$.
Therefore, even though \spin(L) originates in the G-L critical point, it should not be assumed that crossing \spin(L) will produce the G phase.

At the same time, the CPF scenario depicted schematically in Fig.~\ref{cpf}(b) also revives the possibility, implicit in the original predictions of the SLC, that the liquid phase might cavitate to the gas phase on cooling at constant $P>0$, so long as $P$ is not too large.  That is, if the liquid is cooled isobarically at a sufficiently low positive $P$, it will encounter \spin(L) at a point where both the LDL and G phases are observable, and where G is the more stable phase of the two. Such a path is illustrated by the horizontal arrow in Fig.~\ref{cpf}(b).  Depending on the non-equilibrium kinetics of the system as it begins to change phase after crossing \spin(L), it thus may be possible to observe liquid cavitation on cooling at small positive $P$, just as predicted by the SLC.  {\color{black}In practice, we expect this outcome to be unlikely, as it is difficult to imagine a kinetic pathway that would completely miss the metastable LDL phase that lies between the densities of HDL and G.  Nonetheless, the thermodynamic permissibility of this intriguing behavior is worth noting.}

\section{Thermodynamic behavior in the Critical-point-free scenario}
\label{other}

Most current estimates for the location of a LLCP in supercooled water place it at positive $P$, in which case the CPF scenario will not be relevant for understanding water itself~\cite{Gallo:2016fd}.
However, it is possible that other water-like systems may exhibit the CPF phase behavior.  
In particular, a recent {\it ab initio} simulation study of liquid Si finds phase behavior that fits the pattern of the CPF scenario, including a reentrant spinodal~\cite{Zhao:2016dw}.  The CFP scenario in this case is particularly complex in that the LDL-HDL binodal, the TMD line, and multiple spinodals seem to converge in the negative pressure region; see Fig.~4 of Ref.~\cite{Zhao:2016dw}.  

Our results clarify the thermodynamic behavior for the CPF scenario as realized in the EVDW model.  The CPF scenario is usually described as the case where the \coex(LDL-HDL) binodal ends because it intersects \spin(L), and where the LDL-HDL critical point disappears into the unstable region~\cite{Gallo:2016fd}.  
As shown in Figs.~\ref{e14} and \ref{cpf}, the CPF scenario arises in the EVDW model when \coex(LDL-HDL) ends on \spin(LDL), not on \spin(L).  That is, the LDL-HDL binodal ends at a point where LDL becomes unstable, but where HDL remains stable.  The LDL-HDL critical point is not hidden in the unstable region; it has ceased to exist.  Furthermore, the identity of \spin(L) does not change in the low-$T$ region where the LDL-HDL transition occurs, because \spin(L) does not make contact with \coex(LDL-HDL).  At a qualitative level, the CPF scenario realized in Fig.~\ref{e14} corresponds well with the behavior found for liquid Si in Ref.~\cite{Zhao:2016dw}, suggesting that the detailed picture provided by the EVDW model can help interpret the results of such simulations.

As noted in Section~\ref{speedy},
we identify several unusual features in the phase diagrams associated with the CPF scenario.  One of these is the ``open" binodal envelope observed in the case of the LDL-HDL phase transition depicted in Fig.~\ref{e14}(a).  Normally we would expect that the high and low density branches of the binodal curve in the $\rho$-$T$ plane for a liquid-liquid phase transition 
should merge at a critical point as $T$ increases.  Here we see that this merger is pre-empted by the loss of stability of one of the phases, in this case LDL, and the result is an open binodal.  
An implication of this behavior is the existence of a continuous thermodynamic path by which a stable phase [e.g., a HDL state point lying outside of the \coex(LDL-HDL) envelope] can be converted to a metastable phase [e.g., a HDL state point lying between \coex(LDL-HDL) and \spin(L)] without ever crossing the associated binodal curve.  Such a path is not possible when the binodal is a single continuous boundary in the $\rho$-$T$ plane.

The stability field of LDL in Fig.~\ref{e14}(a) is also unusual in that the region of thermodynamic stability occurs inside a spinodal envelope [\spin(LDL)], rather than outside, which is typically the case.  In addition, we find that \spin(LDL) passes through a maximum in the $\rho$-$T$ plane, similar to the maximum formed by \spin(G) and \spin(L) at the G-L critical point.  However, the maximum of \spin(LDL) is not a critical point because the nearby thermodynamic states at the same temperature as the maximum are all unstable.

A recent work by Anisimov and coworkers~\cite{Anisimov:2018ki} also
studies the phase behavior arising from an analytic equation of state that describes both a gas-liquid and a liquid-liquid phase transition, although their formulation is different from the EVDW model and is based on a free energy of mixing of two interconvertible species.  Like the EVDW model, the model in Ref.~\cite{Anisimov:2018ki} is able to generate both the LLCP and CPF scenarios.  However, the detailed topology of the binodals and spinodals in the CPF case is qualitatively different from that found here.  In Ref.~\cite{Anisimov:2018ki}, \spin(L) is always monotonic in the $T$-$P$ plane (i.e. it is never reentrant) and so the implications and viability of the SLC are not tested by their model. 
Also, we find that the TMD line is observed at $P>0$ in both the LLCP and CPF scenarios, while in Ref.~\cite{Anisimov:2018ki} the TMD retreats to negative $P$ and disappears completely when approaching the CPF scenario.  Ref.~\cite{Anisimov:2018ki} does not present the behavior of all of the spinodals in their model, or analyze the intersection of a binodal and a spinodal, and so it would be useful to clarify the differences in phase behavior between the EVDW model and that of Ref.~\cite{Anisimov:2018ki}.  Nonetheless, comparison of Ref.~\cite{Anisimov:2018ki} with our results demonstrates that there are distinct ways to realize the CPF scenario.

Ref.~\cite{Anisimov:2018ki} 
also finds a ``bird's beak" singularity in the shape of \coex(LDL-HDL) when it touches \spin(L), and speculates that this behavior is a thermodynamic requirement; see Fig.~5 of Ref.~\cite{Anisimov:2018ki}.
We find a different behavior.
As shown in Figs.~\ref{e166} and \ref{e165}, and highlighted in Fig.~\ref{newcrit}, we find that 
when $\ehb$ is between $17.42$~kJ/mol and $\ehb^0$,
the close approach of 
\coex(LDL-HDL) to \spin(L) 
generates a second LDL-HDL critical point and binodal.  The two LDL-HDL critical points merge and disappear at $\ehb=\ehb^0$, at which point the LDL stability field becomes completely isolated in the $\rho$-$T$ plane.  
Although we do not expect such exotic behavior to be universal, or even common, our results show that binodal topologies other than a ``bird's beak" are possible at the point of contact of
\coex(LDL-HDL) and \spin(L).

\begin{figure}
\centerline{\includegraphics[scale=0.3]{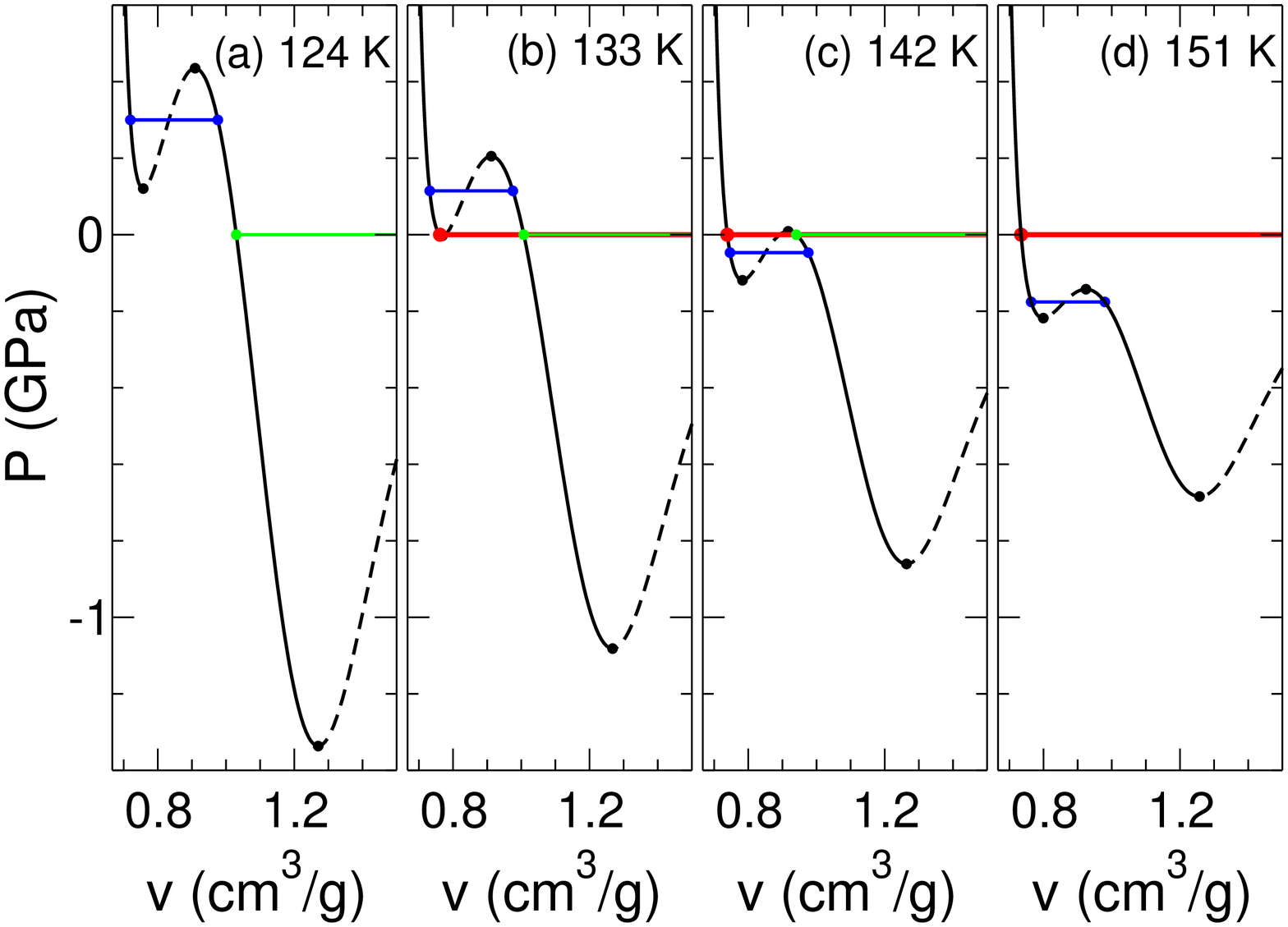}}
\centerline{\includegraphics[scale=0.3]{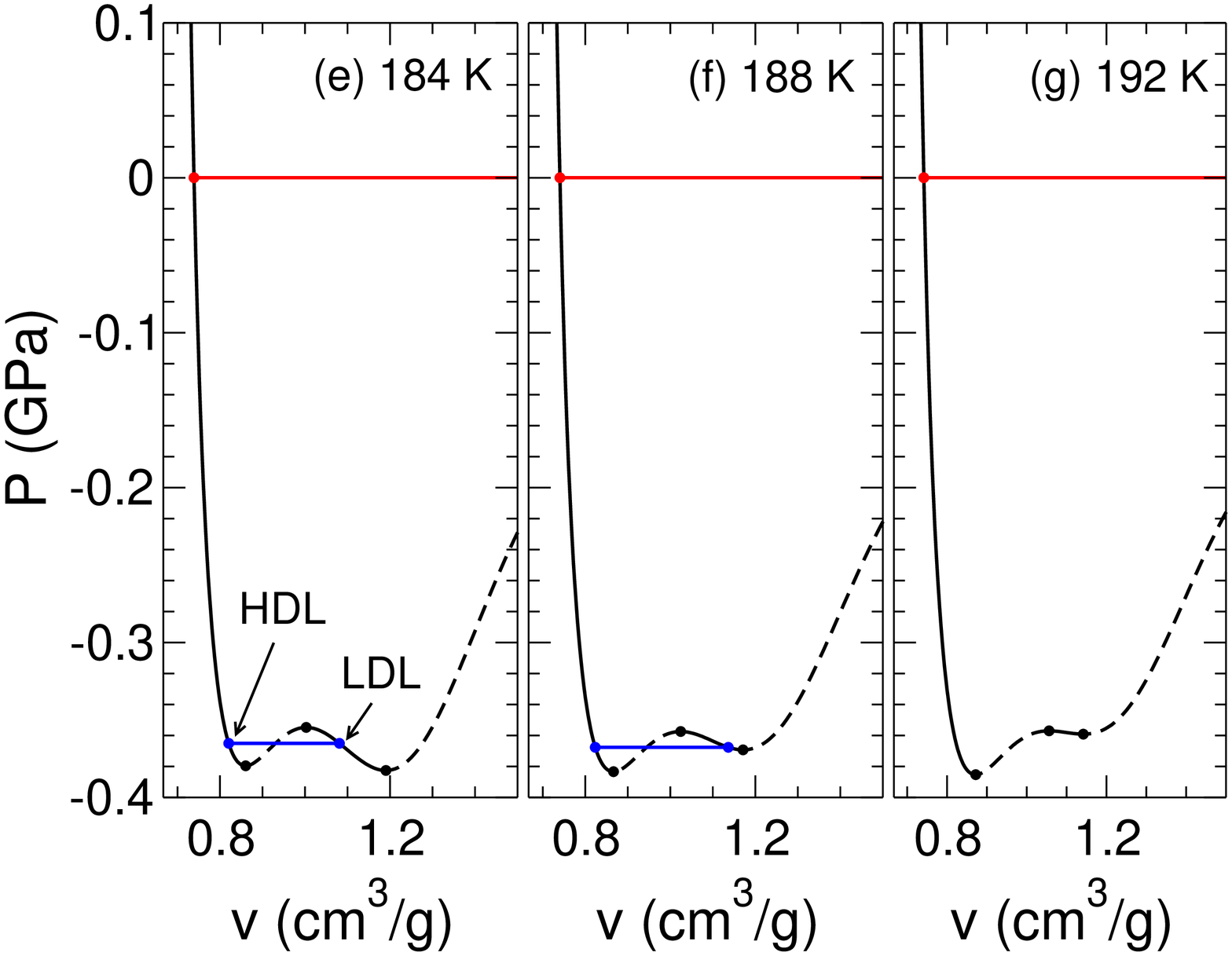}}
\caption{Various $P(v)$ isotherms for $\epsilon_{\rm HB}=-14$~kJ/mol.  
Isotherms are plotted for temperatures (a) 124~K, (b) 133~K, (c) 142~K, (d) 151~K, (e) 184~K, (f) 188~K, and (g) 192~K. 
Black dashed lines identify unstable regions of $P(v)$, while black solid lines are used for stable regions.  Spinodal points are shown as black circles.  Horizontal lines connect coexisting phases as determined by the equal-area construction.  Coexistence of LDL-HDL (blue), G-LDL (green), and G-L (red) are shown.  Note that the stable branch of $P(v)$ for the gas phase is not shown in these plots, as it would occur very far to the right on this scale.}
\label{EOS-1}
\end{figure}

\begin{figure}
\centerline{\includegraphics[scale=0.25]{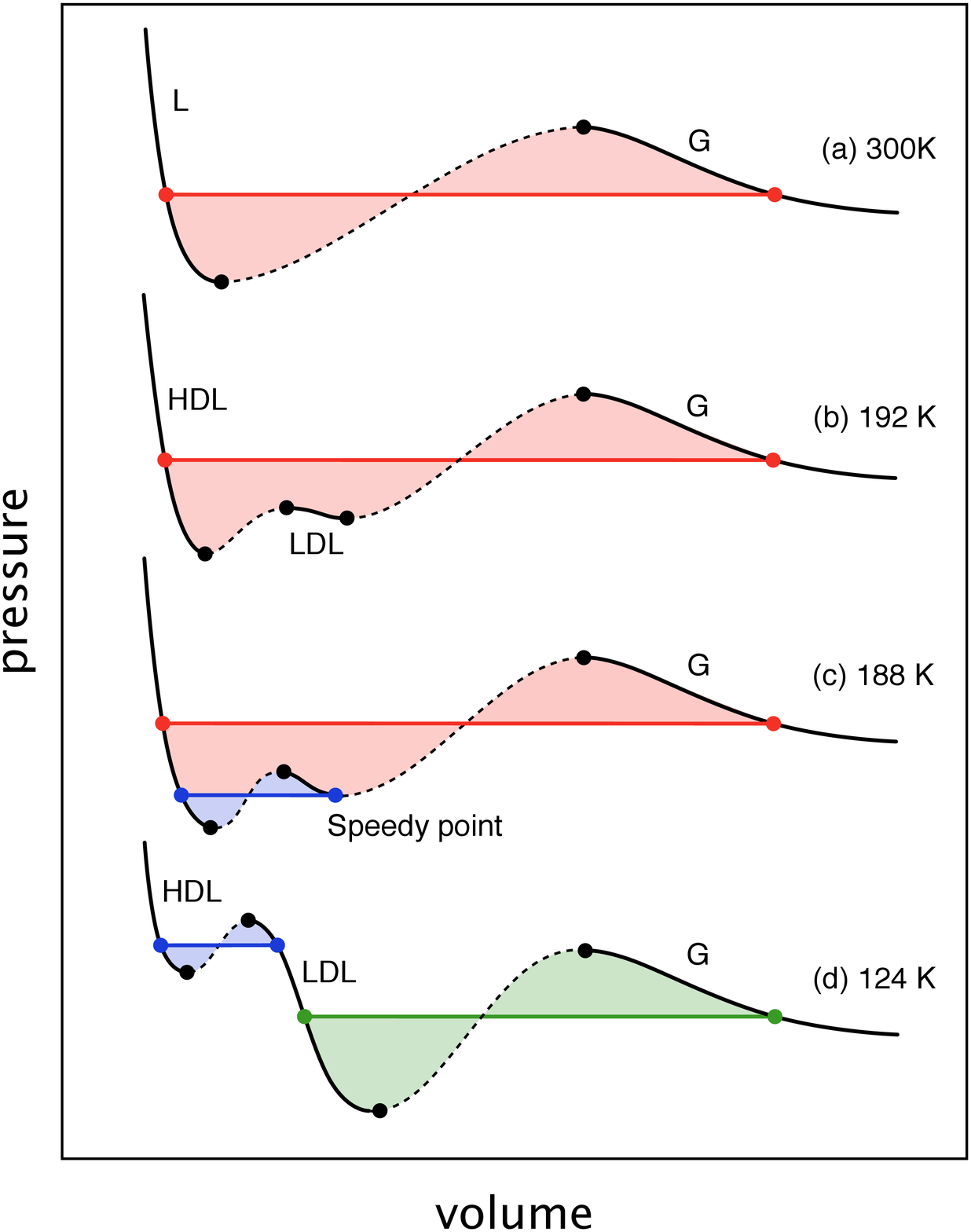}}
   \caption{Schematic $P(v)$ isotherms for the EVDW model 
   for $\epsilon_{\rm HB}=-14$~kJ/mol
      showing the qualitative shape of $P(v)$ at various values of $T$.
      Isotherms are plotted for temperatures (a) 300~K, (b) 192~K, (c) 188~K, and (d) 124~K.
Black dashed lines identify unstable regions of $P(v)$, while black solid lines are used for stable regions.  Spinodal points are shown as black circles.  Horizontal lines connect coexisting phases as determined by the equal-area construction (shaded regions).  Coexistence of LDL-HDL (blue), G-LDL (green), and G-L (red) are shown.  In (c), note that the lower lobe of the LDL-HDL equal-area construction (shaded in blue) also contributes to the area of the lower lobe of the G-L equal-area construction (shaded in red).}
   \label{EOS-2}
\end{figure}

\section{discussion}

Our results demonstrate that the predictions of the SLC can be realized as part of a broader pattern of behavior occurring in a system with both a gas-liquid and liquid-liquid phase transition.  While Debenedetti's critique of the SLC is correct for a system having only a gas-liquid phase transition, we find that the EVDW model provides examples of spinodal-binodal intersections that are not critical points, which we have termed Speedy points.  Our results also clarify the thermodynamic behavior associated with the CPF scenario, which may be relevant for systems such as liquid Si. 

{\color{black}
Our results also show that Speedy points arise naturally when an equation of state possesses more than two distinct density ranges in which the system is thermodynamically stable.
We illustrate this point in Fig.~\ref{EOS-1}, which shows a number of isotherms of $P(v)$ obtained from the EVDW model for $\ehb=-14$~kJ/mol.  Fig.~\ref{EOS-2} 
shows corresponding schematic $P(v)$ isotherms, in order to better describe the qualitative shape of the equation of state over the full range of $v$.
Phase separation requires that $P(v)$ develops a loop, historically called a ``van der Waals loop".  Below the G-L critical temperature, this loop gives rise to a range of $v$ in which the system is thermodynamically unstable [i.e. where $(\partial P/\partial v)_T>0$], bounded by spinodal points [where $(\partial P/\partial v)_T=0$], as shown in Fig.~\ref{EOS-2}(a).
The equal-area construction (equivalent to imposing equal $T$, $P$ and chemical potential for coexisting phases) is then used to determine the properties of the coexisting phases, one on each of the stable branches of $P(v)$ that bracket the unstable region; see the shaded regions in Fig.~\ref{EOS-2}.

As $T$ decreases in the EVDW model, an isolated stable interval appears in $P(v)$ inside the unstable region spanned by the L and G phases; see Fig.~\ref{EOS-1}(g) and Fig.~\ref{EOS-2}(b).  
This stable interval corresponds to LDL, and is bounded by two spinodal points.   However, when it first appears as $T$ decreases, the stable LDL interval of $P(v)$ is not able to form an equal-area construction with either of the other two stable branches of $P(v)$.  Only when $T$ decreases further does it become possible for LDL to form a coexistence with the liquid phase, now termed HDL to distinguish it from LDL.  As shown in 
Fig.~\ref{EOS-1}(f) and Fig.~\ref{EOS-2}(c), the first opportunity on cooling for LDL and HDL to coexist occurs when the low-$P$ LDL spinodal point coincides with the high-$v$ point of the equal-area construction.  This is the Speedy point found at the end of the \coex(LDL\hh HDL) binodal, plotted as a blue open circle in Figs.~\ref{e14} and \ref{cpf}.

As $T$ decreases further, the van der Waals loop associated with the LDL-HDL coexistence moves upward in pressure; see Figs.~\ref{EOS-1}(a-d).  As this loop passes through the pressure range that allows for coexistence with the gas phase (essentially $P=0$ on the scale used in Fig.~\ref{EOS-1}), two new Speedy points occur.  One arises when the high-$P$ LDL spinodal point coincides with the first occurrence on cooling of coexistence between G and LDL, shown in Fig.~\ref{EOS-1}(c).  This is the Speedy point found at the end of the \coex(G\hh LDL) binodal, plotted as a green open circle in Figs.~\ref{e14} and \ref{cpf}.
The final Speedy point occurs when the low-$P$ liquid spinodal point coincides with the last occurrence on cooling of coexistence between the (high-density) liquid and gas, shown in Fig.~\ref{EOS-1}(b).  This is the Speedy point found at the end of the \coex(G\hh L) binodal, plotted as a red open circle in Figs.~\ref{e14} and \ref{cpf}.  At still lower $T$, the van der Waals loops for the LDL-HDL coexistence and the G-LDL coexistence become well separated, as shown in Fig.~\ref{EOS-1}(a) and Fig.~\ref{EOS-2}(d).  
}

Although a Speedy point is not a conventional critical point, it retains some critical-like characteristics.  As noted above, one of the two phases associated with the coexistence will experience diverging fluctuations on approach to the Speedy point, similar to those at a critical point.  However, the other phase remains stable at the Speedy point, with well-defined non-divergent properties.  

Finally, it is important to note that a spinodal is only a well-defined concept in the limit of a mean-field system~\cite{Binder:1987}, which is implicit in our analysis because we study an analytic model equation of state.  In a molecular system (real or simulated), a spinodal limit cannot be approached arbitrarily closely because nucleation processes will transform the metastable phase to a more stable phase before the spinodal is reached~\cite{Binder:2012ey}.
Defined as the intersection of a spinodal and binodal, a Speedy point is also necessarily a mean-field concept, and will not be directly accessible in a realistic system.  However, previous work has demonstrated the value of studying spinodal phenomena in order to place constraints on the range of phase behavior that may be possible in a real system~\cite{Speedy:1976,Speedy:1982fd,Gallo:2016fd,Handle:2017is,Rovigatti:2017fb}.
The present work highlights the importance of considering the thermodynamic phenomena that arise along the metastable extension of a binodal, which are common in any multi-phase system, especially near a triple point.  Speedy points, considered as idealized limits for metastable binodals, may therefore be relevant in a wide range of metastable behavior. 

\begin{acknowledgements}
We dedicate this work to the memory of Robin J. Speedy, whose insights have 
shaped so much of our understanding of metastable water.
PC thanks St.~Francis Xavier University for support.  
FS acknowledges support from MIUR-PRIN  2017Z55KCW.
PHP thanks NSERC and the Dr.~W.F.~James Research Chair Program for support, and C.A. Angell for valuable discussions.  
\end{acknowledgements}

\bibliography{slc}


\end{document}